\def\simlt{\lower.5ex\hbox{$\; \buildrel < \over \sim \;$}}
\def\simgt{\lower.5ex\hbox{$\; \buildrel > \over \sim \;$}}
\RequirePackage{lineno} 
 \documentclass[preprint2]{emulateapj}
 \usepackage{graphicx}
 \usepackage[export]{adjustbox}
 \usepackage{float}

\newcommand{\myemail}{mrline@ucsc.edu}
\slugcomment{Submitted to the Astrophysical Journal }
\shorttitle{Brown Dwarf Retrieval}
\shortauthors{Line et al.}

\begin{document}
\title{Uniform Atmospheric Retrieval Analysis of Ultracool Dwarfs I: Characterizing Benchmarks, Gl570D and HD3651B}
\author{Michael R. Line}
\affil{Department of Astronomy and Astrophysics, University of California, Santa Cruz, CA 95064}
\author{Johanna Teske}
\affil{Carnegie DTM, 5241 Broad Branch Road, NW, Washington, DC 20015}
\affil{Carnegie Origins Fellow, jointly appointed by Carnegie DTM \& Carnegie Observatories}
\author{Ben Burningham}
\affil{NASA Ames Research Center, Mail Stop 245-3; Moffett Field, CA 94035}
\affil{Center for Astrophysics Research, Science and Technology Research Institute, University of Hertfordshire, Hatfield AL10 9AB, UK}
\affil{Marie Curie Fellow}
\author{Jonathan J. Fortney}
\affil{Department of Astronomy and Astrophysics, University of California, Santa Cruz, CA 95064}
\author{Mark S. Marley}
\affil{NASA Ames Research Center, Mail Stop 245-3; Moffett Field, CA 94035}

% going to follow "group leader goes last" tradition

\altaffiltext{1}{Correspondence to be directed to \myemail}

\begin{abstract}

Interpreting the spectra of brown dwarfs is key to determining the fundamental physical and chemical processes occurring in their atmospheres.  Powerful Bayesian atmospheric retrieval tools have recently been applied to both exoplanet and brown dwarf spectra to tease out the thermal structures and molecular abundances to understand those processes.  In this manuscript we develop a significantly upgraded retrieval method and apply it to the SpeX spectral library data of two benchmark late T-dwarfs, Gl570D and HD3651B, to establish the validity of our upgraded forward model parameterization and Bayesian estimator. Our retrieved metallicities, gravities, and effective temperatures are consistent with the metallicity and presumed ages of the systems. We add the carbon-to-oxygen ratio as a new dimension to benchmark systems and find good agreement between carbon-to-oxygen ratios derived in the brown dwarfs and the host stars. Furthermore, we have for the first time unambiguously determined the presence of ammonia in the low-resolution spectra of these two late T-dwarfs.   We also show that the retrieved results are not significantly impacted by the possible presence of clouds, though some quantities are significantly impacted by uncertainties in photometry.   This investigation represents a watershed study in establishing the utility of atmospheric retrieval approaches on brown dwarf spectra.

\end{abstract}

\section{Introduction}

The spectrum of a brown dwarf opens a series of windows into the depths of its atmosphere, revealing its composition and thermal structure. Differing wavelengths peer to a diversity of depths and are influenced by varying atmospheric species. Teasing out   the chemical composition and atmospheric structure of brown dwarfs by synthesizing the information emerging from each window
is key to understanding the processes acting in their atmospheres and possibly their evolutionary histories. Historically, empirical analysis techniques, such as color magnitude diagrams, the appearance of spectroscopic features, or comparisons
to forward models have been used to understand broad trends in the brown dwarf population.  Such analyses have led to the L-T-Y classification schemes as well as widely accepted spectral and photometric indicators  of gravity and metallicity (Kirkpatrick 2005; Cushing et al. 2011). However such approaches have important limitations. The physics and chemistry of these dense molecular atmospheres is complex and subtle
processes can significantly influcence thermal spectra. 

The traditional method for interpreting brown dwarf spectra is to construct sets of sophisticated  coupled radiative-convective-chemical equilibrium models to which the data can be compared (Marley et al. 1996; Allard et al. 1996; Tsuji et al. 1996; Burrows et al. 2001).  Systematic
comparisons of models to data constrain atmospheric composition, temperature, and cloud properties (e.g., Saumon et al. 2006; Cushing et al. 2008; Rice et al. 2010).
While these grid modeling studies have unquestionably advanced our understanding of brown dwarf atmospheres, there is still 
much that we don't understand.  For instance, can dynamical processes in brown dwarf atmospheres cause deviations from radiative convective equilibrium?  How do their molecular compositions vary with altitude?  Can we measure the molecular abundances rather than
 assume them based upon chemical equilibrium or disequilibrium?  Can their elemental abundances deviate from the often assumed solar composition? It is difficult to answer such questions from comparison solely with forward models, as not all processes can be modeled with fidelity and, in any case, it is not always obvious which processes are responsible for given deviations of models from data.

Line et al. (2014b) presented a novel approach for answering the above questions.  Rather then rely on grid models that only constrain a few basic parameters, they used atmospheric retrieval methods common in Earth and planetary atmosphere studies to invert brown dwarf spectra for the temperature structures and abundances.  Using such approaches allows the maximum amount of information to be extracted from brown dwarf spectra.  Line et al. (2014b) found for, Gl570D, with few assumptions about the nature of the chemistry or the temperature-pressure profile, that the retrieved quantities were consistent with previous grid modeling studies (Saumon et al. 2006), though with some minor deviations such as a more isothermal upper atmosphere and a depleted ammonia abundance.  While individual objects themselves are interesting, the key to understanding the underlying physical processes in brown dwarf atmospheres (or any class of atmospheres) requires an investigation of a large population of objects.  With large populations one can identify trends and correlations of various physical parameters that can lead to insight into new phenomena. 
%(e.g., why some exoplanets do or don't have thermal inversions, Knutson et al. 2010).  

%add how studying benchmark systems can validate retrieval

We first aim to improve upon the techniques presented in Line et al. (2014a) and to validate our improved approach against benchmark brown dwarfs.  Then in a followup paper we will apply our new approach to a small sample of T-dwarfs.  In that sample we will identify trends within our retrieved results such as metallicity vs. gravity, molecular abundances vs. effective temperature etc., and also how empirical spectral properties such as color or other indices correlate with the retrieved physical parameters.  Preliminarily, we choose mid- to late-T dwarfs as they, based upon previous investigations, appear to be largely free of thick silicate clouds that can complicate the interpretation of L and early-T spectra (Kirpatrick 2005).  

In this manuscript we present the upgraded retrieval approach and apply it on two benchmark brown dwarfs, HD3651B and Gl570D, using the the SpeX Prism Library (Burgasser et al. 2006a).  In $\S$\ref{sec:Methods} we present a modified Bayesian retrieval approach and a novel approach for inverting for temperature structures.  In $\S$\ref{sec:Results} we present our retrieved results and how they are impacted by various assumptions.  Finally, in $\S$\ref{sec:benchmark} we compare our retrieved values to benchmark properties such as the age and metallicity. Additionally, we present a detailed stellar abundance analysis of Gl570A and HD3651A and derive their carbon-to-oxygen ratios in order to add another dimension to the benchmark comparison.

\section{Methods}\label{sec:Methods}
Here we give a brief review of current state of knowledge of Gl570D and HD651B,  describe the data we use, the forward radiative transfer model, and the inverse methods to retrieve temperature and abundance information. Building upon Line et al. (2014b), we have made significant upgrades to our methodology in terms of the forward model, Bayesian estimator, and treatment of the data.  We highlight those differences where applicable. 

\subsection{Current State of Knowledge on Gl570D \& HD3651B}
The targets of this study have been selected for their suitability as robust test cases for validating our approach. Both objects are wide-orbit common proper motion companions to stars, allowing us to check our derived properties for consistency with those found for their stellar hosts by other routes. For this reason, such systems are often referred to as benchmarks, although the quality and context of the available constraints varies widely depending the mass and evolutionary phase of the stellar primary (e.g., Pinfield et al. 2006).

Gl570D was the first T-dwarf companion to a star identified following the prototypical T-dwarf Gl229B (Burgasser et al. 2006a). It is a wide component in a hierarchical quadruple system, whose inner components are an M1V+M3V spectroscopic binary (Gl570B and C) and a K4V primary (Gl570A) (Gleise 1969; Duquennory  \& Mayor 1988; Mariotti et al. 1990; Foreville et al. 1999), from which Gl570D lies at a projected separation of $1525 \pm 25$~AU (Burgasser et al. 2000).
Gl570D has been subject to a number of grid-model fitting studies, which have to varying extents used the primary star to restrict the parameter space available for the models (e.g., Geballe et al. 2001; Saumon et al. 2006; Legett et al. 2007; Saumon et al. 2012) and has been used as an anchor point for applying trends seen in self-consistent grid models to estimate parameters of the wider T~dwarf population (Burgasser et al. 2006b).

Liu, Leggett \& Chiu (2007) used a variety of stellar age indicators for Gl570A to constrain the age of the system to the range 1--5~Gyr, whilst Saumon et a. (2006) collated literature values to estimate its metallicity as ${\rm [Fe/H]} = 0.09 \pm 0.04$. In the Appendix we present a new measurement of ${\rm [Fe/H]} = -0.05\pm 0.17$.

HD3651B was identified as a wide common proper motion companion to the planet-hosting K0V star HD3651A with a projected separation of 480~AU (Murgrauer et al. 2006; Luhman et al. 2007; Liu, Leggett \& Chiu 2007).
Like Gl570D, it has been the subject of a number of spectroscopic studies that have been constrained by the properties of the primary star (e.g., Leggett et al. 2007; Burgasser 2007). 
Liu, Leggett \& Chiu (2007) reviewed X-ray luminosity, chromospheric and rotation-based age indicators for HD3651A, and found the target lies in the unreliable ``older" tail of each of these diagnostics. Like them, we adopt the isochronal age range of 3--12Gyrs from Velenti \& Fischer (2005).
As an exoplanet host star, HD3651A has been the subject of several recent composition studies. The determinations of ${\rm [Fe/H]}$ are consistently super-Solar, ranging from ${\rm [Fe/H]} = +0.12 \pm 0.04$ (Santos et al. 2004) to ${\rm [Fe/H]} = +0.19 \pm 0.03$ (Ghezzi et al. 2010).  In the Appendix we have further developed these targets potential to contribute to our understanding its substellar companions by compiling detailed abundance measurements from the literature, and measuring new abundances for both.  Most significantly for this study we present new determinations of the C to O ratios for both primary stars.

%Abundances have also been measured for a number of other species by (Delgado-Mena et al.  2010; Allende-Prieto et al. 2004); though the bulk of these are not important for our current study, we highlight their measurements of C/O = 1.02$\pm 0.1$ and C/O=0.59  due to its relevance to our analysis (see Appendix for caveats on these measurements). However, followup investigations (Fortney 2012; Teske et al. 2013; Nissen 2013) suggest that the C/O measurements in Delgado-Mena et al.  (2010) are systematically overestimated.  We discuss more on the stellar carbon-to-oxygen measurements in the appendix.

\subsection{Data}
%Why Gl570D and HD3651B?
We use the data within the SpeX Prism Library (Burgasser et al. 2006) to perform our analysis on our target objects.  Since a given SpeX spectrum is continuous, we avoid having to consider various instrumental systematics (e.g., as had to be done in Line et al. 2014b) and subsequent impact on their interpretation that come along with having to stitch spectra from multiple instruments together.   We use the SpeX data taken in the SXD mode which cover wavelengths between 0.8 - 2.5 $\mu$m with a wavelength dependent resolving power ranging from 87 to 300.  The spectra within the SpeX SXD library are oversampled relative to a spectral resolution element, and are therefore each pixel is not an independent sample. Using the full oversampled data would result in over constrained results.  We therefore sample every few pixels.  We choose the sampling length based upon the auto-corrletion length scale of the residuals of a typical model fit to the data. This is 2.7 pixels. We thus take every other 3rd pixel to be statistically independent. An alternative approach would be to model the covariant error structure of the oversampled data through a Guassian process (e.g., Czlecka et al. 2014).  

\begin{figure}
\begin{center}
\includegraphics[width=0.5\textwidth, angle=0]{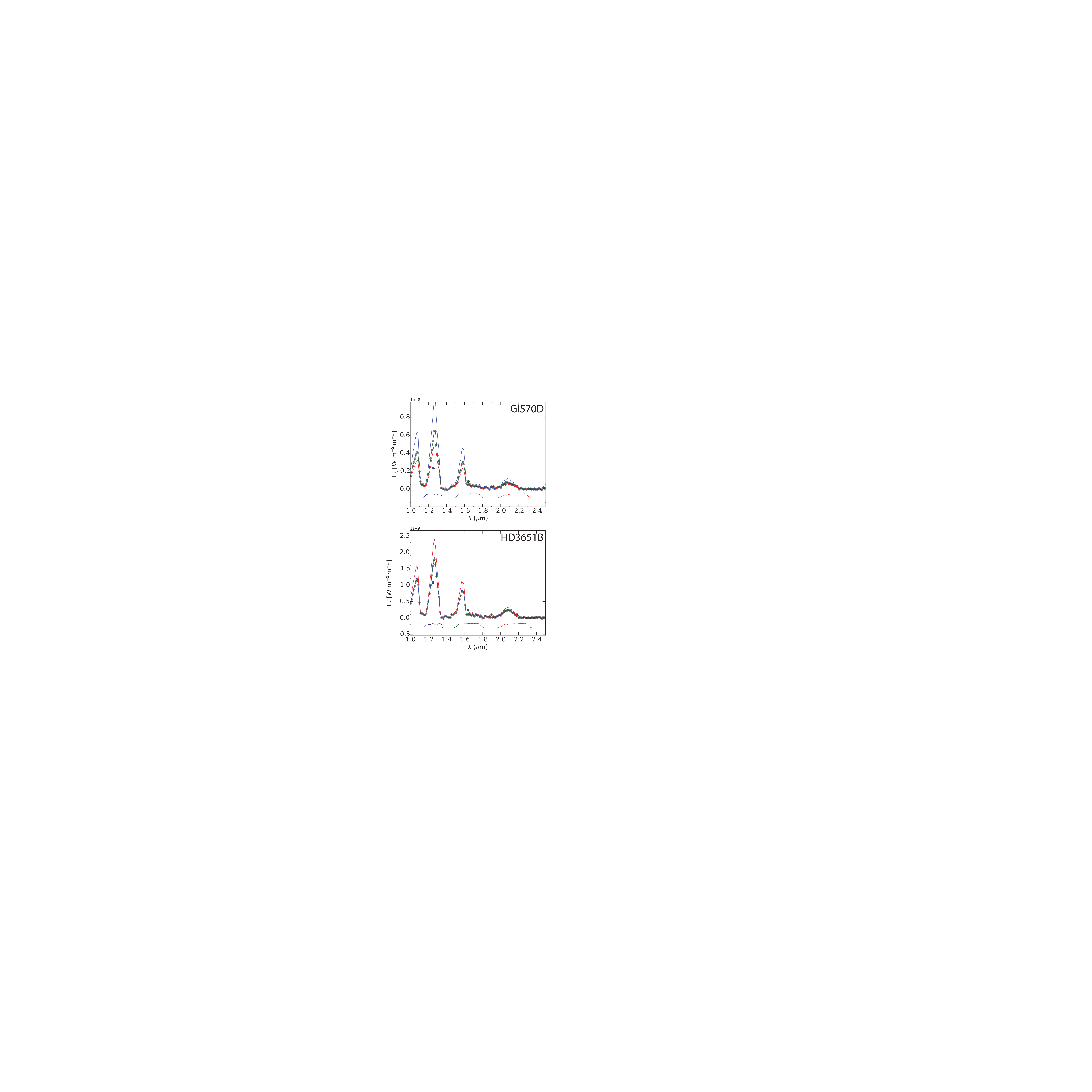}
\end{center}
     \caption{ \label{fig:figure1} Spectral Calibration Process on the two objects.  The photometric (circles) J (blue), H (green), and K (red) points are used to calibrate the normalized SpeX spectra.  The spectra are calibrated by integrating over the corresponding filter profile and rescaling the normalized spectrum to match the photometric fluxes.  Each photometric point gives a different calibrated spectrum--blue for J-band, green for H-band, and red for K-band.  The H-band calibrated spectrum with the error bars is shown with gray diamonds and is what we use to perform our analysis.  Note that the differences between each band integrated spectrum greatly exceeds that of the photometric uncertainty. }
  \end{figure}

The native format of the spectral fluxes and error bars within the library are normalized spectra.  In order to perform the subsequent analysis, these spectra and error bars must be converted into physical units via photometric calibration.  The SpeX database provides the 2MASS photometric J, H, and K-band magnitudes for each object.  We convert the 2MASS magnitudes into MKS flux units (W m$^2$ m$^{-1}$) using the Spitzer Science Center Magnitude/Flux density converter \footnote{http://ssc.spitzer.caltech.edu/warmmission \\/propkit/pet/magtojy} which uses the zero point fluxes described in Cohen, Wheaten \& Megeath (2003) . The H-band flux is used to derive the final flux-calibrated spectrum just as in Saumon et al. (2006) and Liu, Leggett \& Chiu (2007).  Figure \ref{fig:figure1} shows the flux calibrated spectra.  Unfortunately, the calibrated spectra are different by some scale factor that is larger than the quoted photometric uncertainty, depending upon which photometric point is used to calibrate.  Our retrieval model (discussed below) includes a scaling factor as a free parameter so these differences are not critical to our analysis (with the exception of the derived spectroscopic radius, see the Results section for more on this).  We do note, however, that better photometry, namely (MKO) exists for these objects.  Furthermore, Stephens \& Leggett (2004) suggest that the 2MASS photometry is not the most accurate due to the shape of the 2MASS filter profiles with respect to telluric transmittance. They also provide correction factors for the 2MASS photometry based upon the more accurate MKO photometry. However, we choose to use the un-corrected 2MASS photometry as they shall provide the most conservative impact of inconsistencies in photometry on the derived quantities.
 
%($http://ssc.spitzer.caltech.edu/warmmission/propkit/pet/magtojy/#mag_to_fnu$) 

%%%%%%%%%%%%figure1%%%%%%%%%%%%%%%%%%%%Spectral Calibration
%%%%%%%%%%%%figure1%%%%%%%%%%%%%%%%%%%%

\subsection{Forward Radiative Transfer Model}\label{sec:FM}
The forward radiative model is a derivative of the CHIMERA forward model (Line et al. 2013a; 2014a,b) which computes the upwelling 1-dimensional disk integrated thermal emission spectrum given the molecular abundances, temperature-pressure (TP) profile, and gravity  $g$.  Near infrared spectra of late-T's are typically dominated by strong absorption features from water, methane, and alkali metals and little if any obvious absorption exists due to other gases.  We include constant-with-altitude volume (molar) mixing ratios for H$_2$O, CH$_4$, CO, CO$_2$, NH$_3$, H$_2$S, and alkali opacities.  These are the species known to be found in cool dwarf atmospheres that have spectral signatures in the near infrared.  The alkali opacities include only sodium and potassium and are treated as only one free parameter with their ratio assumed to be solar.  Hydrogen/helium in solar ratio is assumed to make up the remainder of the gas.   

The TP profile is also included as a free parameter (see \S \ref{sec:TP_prior}).  The TP profile is partitioned into 15 evenly spaced slabs (or knots) in log pressure between 315 bar and 1 mbar.  Because a 15 layer atmosphere does not have enough vertical resolution to accurately compute fluxes, the 15 level profile is spline interpolated to a finer 70 level grid before before being passed to the radiative transfer.   Using fewer TP points permits swifter convergence of the Bayesian estimator.

In addition to the TP profile and gas mixing ratios, we include gravity as a free parameter.  Gravity controls the column optical depth.  Most of the opacity database is drawn from Freedman et al. (2014) and references therein as in Line et al. (2014b), but we have also incorporated the most up-to-date methane line list (Yurchenko et al. 2014) using the line broadening coefficients from Margolis (1996).  Instead of using line-by-line or correlated-K we simply sample the hi-resolution cross sections at 1 cm$^{-1}$ resolution (see Sharp \& Burrows 2007 section 2) which is more than sufficient for moderate resolution spectra. 

Finally, the high resolution spectra are convolved with a wavelength dependent Gaussian instrumental profile that reflects the wavelength dependent resolving power, and then binned to the data wavelength grid in order for direct data-model comparison.  

\begin{figure}
\begin{center}
\includegraphics[width=0.45\textwidth, angle=0]{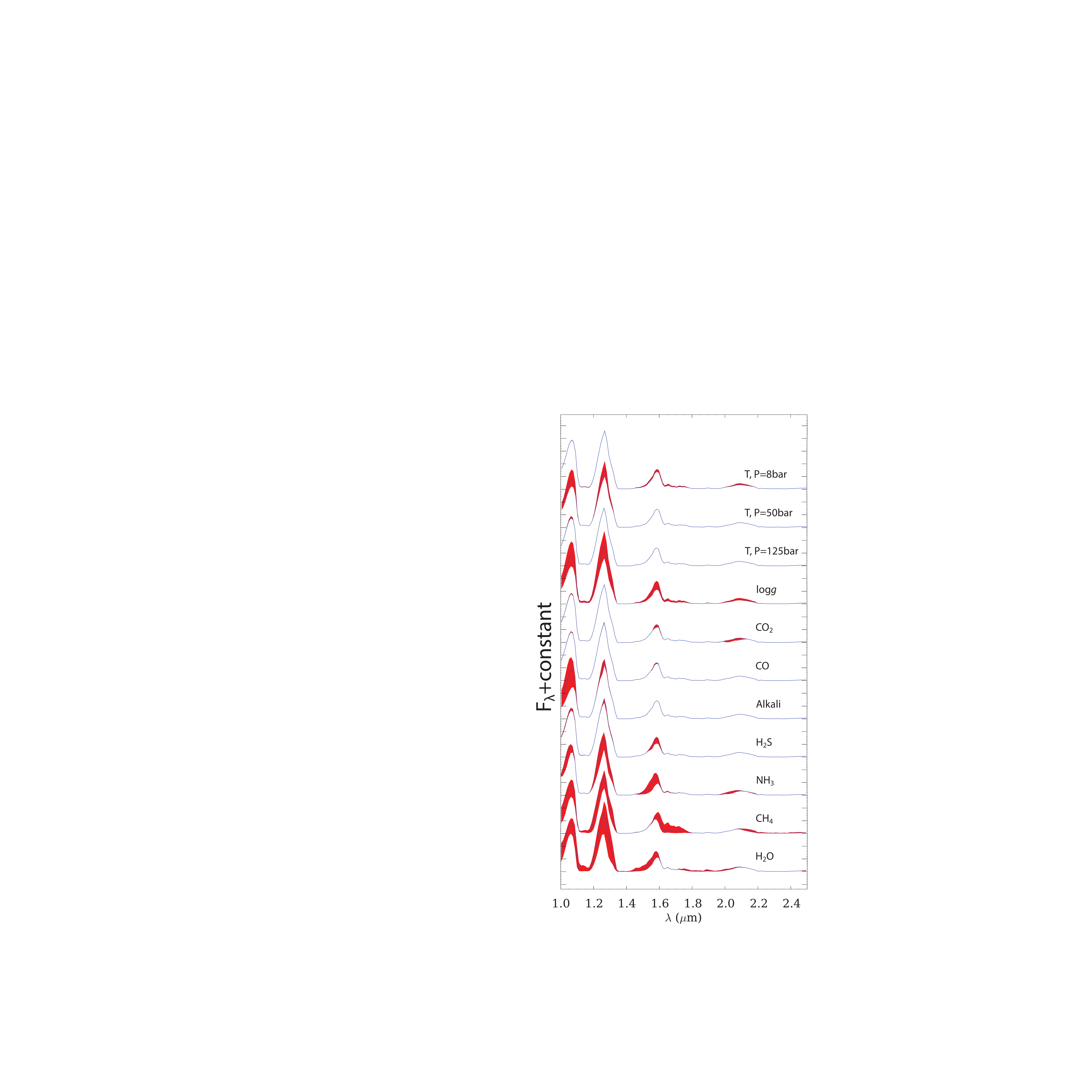}
\end{center}
     \caption{ \label{fig:figure2} Sensitivity of the spectrum to various parameters.  This is a synthetic spectrum with an effective temperature (equivalent black body flux integrated over the whole spectrum) of 700 K and a log$g$ of 5 and with purely thermochemical equilibrium composition.  The red regions represent a change in the spectrum due to a perturbation of each of the parameters.  For H$_2$O, CH$_4$,  NH$_3$, H$_2$S, alkali, this perturbation is $\pm$ 0.5 dex (where 1 dex is one increment in log space, or 1 order of magnitude) in number mixing ratio from the thermochemical abundance value. For CO the perturbation is +2 dex, CO$_2$, +6 dex, log$g$$\pm$ 0.1 dex, and the temperatures are perturbed at each level by $\pm$50 K.    }
  \end{figure}

Figure \ref{fig:figure2} shows the sensitivity of a model spectrum at SpeX resolutions to the various parameters.    Many of these parameters are sensitive to similar wavelengths.  This will result in correlations/degeneracies amongst the gases and the temperature at different levels in the atmosphere.  

All of the aforementioned parameters (23 of them) control the flux at the top of a brown dwarf atmosphere.   We also include additional ``systematic" parameters that facilitate the direct comparison of the model to the data.  These parameters account for the radius to distance ratio (and implicitly the flux calibration), uncertainties in the wavelength calibration, underestimation of the spectral error bars, and a smoothing parameter for the temperature profile-for a total of 27 free parameters.  These ``systematic" parameters will be discussed in more detail in the following section.  A list of all of the parameters and a brief description of each is presented in Table \ref{tab:table1}. The model has many parameters, but this allows us to make very few implicit assumptions about the nature of the atmosphere.  Unconstrained parameters will simply appear unconstrained.  The beauty of modern retrieval approaches (below) is that they can accommodate numerous parameters and fully account for all of the correlations amongst them.  A larger number of parameters will of course result in more conservative estimate of the uncertainties through marginalization.  

%%%%%%%%%%%%%%%%%%%%%%%%%%Table1-Parameters%%%%%%%%%%%%%%%%%%%%
\begin{table}
\centering
\caption{\label{tab:table1} Parameters in the forward model. }
%\resizebox{\textwidth}{!}{%
\begin{tabular}{cc}
%\footnote{\tiny{aa}}
\hline
\hline
\cline{1-2}
Parameter & Description    \\
\hline
log$f_{i}$ & log of the  volume mixing ratios of H$_2$O, \\ 
   & CH$_4$, CO, CO$_2$, NH$_3$, H$_2$S, and alkali (Na+K)  \\
   log$g$ & log gravity [cms$^{-2}$]\\
   (R/D)$^2$ & radius-to-distance scaling (R$_{J}$/pc)\\
  T$_{j}$ & temperature at 15 pressure levels (K)\\
  $\Delta \lambda$ &  uncertainty in wavelength calibration (nm) \\
  $b$ & error bar inflation exponent (equation \ref{eq:eq3})\\
  $\gamma$ & TP profile smoothness hyperparameter\\
\hline
\end{tabular}
\end{table}
%%%%%%%%%%%%%%%%%%%%%%%%%%Table1%%%%%%%%%%%%%%%%%%%%%%%%%%%

%%%%%%%%%%%%%figure2-Sensitivity%%%%%%%%%%%%%%%%%%%%
%%%%%%%%%%%%figure2%%%%%%%%%%%%%%%%%%%%

\subsection{Retrieval Model}
\subsubsection{Bayesian Implementation}
The retrieval model is the Bayesian engine that optimizes the forward model to fit the data.  We use the Markov chain Monte Carlo approach implemented with affine-invariant ensemble sampler, EMCEE (Foreman-Mackey et al. 2013). This is a significant advancement over the optimal estimation and bootstrap Monte Carlo approaches used in Line et al. (2014b) as we are now able to make fewer a priori assumptions about the smoothness of the temperature profile or the Gaussian shape of the parameter uncertainties.  EMCEE requires only a functional form for the log of the posterior probability to perform the optimization.  The posterior probability is a combination of the likelihood and the prior described as follows.  Starting from Bayes theorem 
\begin{equation}
	 p({\bf x|y})=\frac{\mathcal{L({\bf y|x})}p({\bf x})}{E}
\end{equation}
where {\bf x} is the parameter vector described in $\S$\ref{sec:FM} and {\bf y} is the data vector--in our case the spectrum, $p({\bf x|y})$ is the posterior probability distribution, $\mathcal{L({\bf y|x})}$ is the likelihood distribution which penalizes poor fits to the data, $p({\bf x})$ is the prior which represents any external constraints, and $\it{E}$ is a normalization factor known as the evidence, or marginal likelihood, which is required for Bayesian model comparison but not for parameter estimation.  We use the following log-likelihood function:
\begin{equation}\label{eq:lnlike}
	 \ln\mathcal{L({\bf y|x)}}=-\frac{1}{2}\sum\limits_{i=1}^n \frac{(y_{i}-F_{i}({\bf x}))^2}{s_{i}^2}-\frac{1}{2}\ln(2\pi s_{i}^2)
\end{equation}
Here, the index $\it{i}$ denotes the $\it{i^{th}}$ data point, in our case some property at a single wavelength bin,  $\it{y}$ is the measured flux, $\it{F({\bf x})}$ is the modeled flux that comes out of the forward model ($\S$\ref{sec:FM}), and $\it{s}$ is the data error given by
\begin{equation}\label{eq:eq3}
	 s_{i}^2=\sigma_{i}^2+10^{b}
\end{equation}
where $\sigma$ is the measured error for the $\it{i^{th}}$ data point and $\it{b}$ is a free parameter.  Differing from Line et al. (2014b), we modify the standard error on the data point by the factor $10^{b}$ to account for underestimated uncertainties and/or unknown missing forward model physics (Foreman-Mackey et al. 2013, Hogg et al. 2010, Tremain et al. 2002), e.g., imperfect fits. This results in a more generous estimate of the parameter uncertainties.  Note that this is similar to inflating the error bars post-facto in order to achieve reduced chi-squares of unity, except that this approach is more formal because uncertainties in this parameter are properly marginalized into the other relevant parameters.  Generally, the factor $10^{b}$ takes on values that fall between the minimum and maximum of the square of the data uncertainties.  The first term inside the summation in equation \ref{eq:lnlike} is the familiar ``chi-square".  This term penalizes large residuals.  The second term in the summation is the Gaussian normalization factor that is normally excluded from standard fitting routines due to the unchanging data errors.  Because the data errors include the free parameter $b$ this normalization can change, and hence has to be taken into account.  Really, the purpose of this term is to provide a balance for the error bar inflation parameter to prevent it from approaching infinity.  

The prior, $p({\bf x})$, can be broken up into several pieces as
\begin{equation}
	 p({\bf x|y})=p({\bf T})p({\bf x'})p(\gamma)
\end{equation}
where $p({\bf x'})$ is the prior on the log of the gas mixing ratios, the instrumental parameters, gravity, and the radius-to-distance scaling while $p({\bf T})$ and  $p(\gamma)$ are the temperature profile priors. The parameter $\gamma$ is the relative weighting of the temperature prior (see equation \ref{eq:TP_prior}) .   The prior details are shown in Table \ref{tab:table2}.  

Because we have both measured fluxes and parallaxes for each object, we are able to calculate the ``photometric" radius.  If we can measure both radius and a gravity we can then constrain the mass.  We know for brown dwarfs the mass cannot exceed $\sim75-80 M_{J}$ (e.g., Burrows et al. 2001).   Therefore, rather than 
place individual priors on the radius and gravity we enforce a prior on the derived mass to fall between the physical plausible values of 1 and 80 $M_{J}$.  This constraint prevents the retrieved radii and gravities from entering an unphysical region of parameter space.

\subsubsection{A Novel Temperature-Pressure Profile Retrieval Approach}\label{sec:TP_prior}
We present our novel method for retrieving temperature profiles in atmospheres.  A common issue in planetary atmospheric retrievals is how to parameterize the temperature profile in an atmosphere.  There are two philosophies.  One philosophy, mostly used in the exoplanet atmosphere community when the data is sparse, is 
to parameterize the atmosphere with some analytic function that can be described by a small set of parameters (e.g., Madhusudhan \& Seager 2009; Line et al. 2012; 2013 Benneke \& Seager 2012).  This is advantageous as the entire TP structure can be controlled by just a few simple parameters.  It is disadvantageous because it is relying on a $\it{parameterization}$ to infer the temperature 
structure.  This could potentially result in biases in the retrieved profiles.  For instance, if one were to use the simple Eddington approximation for the temperature profile, there would be one free parameter--the mean opacity.  While just one free parameter is ideal, 
we know that the Eddington approximation is a poor approximation for brown dwarf atmospheres because the true opacity is not 
constant with pressure, nor is it gray.  While parameterizations are appealing in their simplicity, they can often times be too much of an oversimplification of the physics, and are thus not appropriate.

The classic planetary science approach, the other extreme, is to retrieve the temperature at each model layer in the atmosphere (e.g., Rodgers 2000; Irwin et al. 2008; Lee et al. 2012) within an optimal estimation framework.  For typical model atmosphere grids this can be anywhere between 50-100 independent temperature-pressure points. This was the approach used in Line et al. (2014b).  Because many of the atmospheric levels are degenerate, the retrieved profiles often result in unphysical oscillations, or ringing (Rogers 2000).  The standard remedy to this problem is to implement some $a$ $priori$ covariance matrix, or a smoothing kernel (or Tikonov Regularization), given some set smoothing length scale and $a$ $priori$ width (Irwin et al. 2008).  While this reduces wild oscillations, this smoothing length scale and width must be chosen $a$ $priori$ and cannot change during the course of a retrieval.  Furthermore, these values are often case specific and must be tuned by hand.  Our novel approach remedies all of the aforementioned issues, and can be readily implemented within a Bayesian framework.  

We borrow much of this work from the non-parametric regression literature (Lang \& Brezger 2004; Rahman 2005; Jullion \& Lambert 2007).  The goal is to allow flexibility to fit for each of the independent temperature-pressure points while preserving smoothness, without having to a priori set the degree of smoothness.    The best way to do that is by penalizing the second derivative of the temperature structure.  The second derivative is the ``roughness" of a function.    Temperature vectors with wild, unphysical oscillations, will have large summed second derivatives.  These types of roughness-penalized non-parametric polynomial fits are known as P-splines.  The degree to which this roughness is penalized is included as a free parameter ($\gamma$).  This variable smoothing is implemented as,  
\begin{equation}\label{eq:TP_prior}
	 \ln p({\bf T})=-\frac{1}{2\gamma}\sum\limits_{i=1}^N (T_{i+1}-2T_{i}+T_{i-1})^2-\frac{1}{2}\ln(2\pi \gamma)
\end{equation}
Inside the sum is the discrete second derivative of the temperature profile at each level, $i$ weighted by $\gamma$.  Based on experimentation (Lang \& Brezger 2004; Rahman 2005; Jullion \& Lambert 2007) the hyperprior on $\gamma$ should take the form of an inverse gamma distribution with the properties shown in Table \ref{tab:table2}.  A variable $\gamma$ allows the data to dictate the degree of smoothing.  If the data warrants little smoothing and there truly are oscillations in the TP profile, then $\gamma$ will be large, lending little weight to the smoothing prior.  When the data does not justify rough TP profiles, $\gamma$ will be small resulting in a larger penalty to rough profiles.  As mentioned in \S \ref{sec:FM},  15-knots, or anchor points are used in our TP profile.  This number is somewhat arbitrary as the number and location of the knots should not matter within this framework just so long as there are enough evenly spaced knots to sufficiently resolve potential structure (Eilers \& Marx 1996).    We have tested higher numbers of knots (30 vs. 15) and have indeed found little sensitivity to the choice.  

The combined log-likelihood and log-prior are readily implemented within the EMCEE sampler.  The MCMC is initialized with a tight Gaussian ball with 8 chains or walkers (Foreman-Mackey et al. 2013) per parameter about an educated starting guess in order to minimize burn in time (the time it takes for the MCMC sampler to locate the posterior distribution).  We note that the final results are insensitive to the initial starting point; poor starting points result in longer burn in times.  Convergence is monitored with the Gelman-Rubin statistic.  The statistics are summarized by drawing thousands of random points from the last $\sim5-10\%$ (which we take to be the posterior) of the ensemble of chains which generally entail 500-1000 independent samples (as dictated by the autocorrelation length scale). In the following sections we describe the physical properties of the two benchmark late T-dwarfs evaluated within this framework.

%%%%%%%%%%%%%%%%%%%%%%%%%%Table2-Priors%%%%%%%%%%%%%%%%%%%%
\begin{table}
\centering
\caption{\label{tab:table2} Summary of priors for each of the parameters.}
%\resizebox{\textwidth}{!}{%
\begin{tabular}{ccc}
%\footnote{\tiny{aa}}
\hline
\hline
\cline{1-2}
Parameter & Prior    \\
\hline
$\log f_{i}$ & uniform-in-log\footnote{We note that uniform-in-log priors may cause issues for a larger parameter set. For larger parameter sets a Dirichlet prior should be used } with $\log f_{i}$ $\ge -12$, $\Sigma_{i}f_{i}\le1$\\
(R/D)${^2}$, $\log g$ & uniform, constrained by $1\le gR^2/G \le 80 M_{\rm J}$ \\
$T_{i}$ & see equation \ref{eq:TP_prior}\\
$\Delta \lambda$ &  uniform (-10 - 10 nm)\\
$b$ & uniform, $0.01\times min(\sigma_i^2) \le 10^{b} \le 100\times max(\sigma_i^2)$  \\
$\gamma$ & Inverse Gamma ($\widetilde{\Gamma} (\gamma;\alpha,\beta)$), $\alpha=1,\beta=5\times10^{-5}$\\
\hline
\end{tabular}
\end{table}
%%%%%%%%%%%%%%%%%%%%%%%%%%Table2-Priors%%%%%%%%%%%%%%%%%%%%

\section{Retrieval Results}\label{sec:Results}
\begin{figure*}
%\begin{center}
\includegraphics[width=1\textwidth, angle=0]{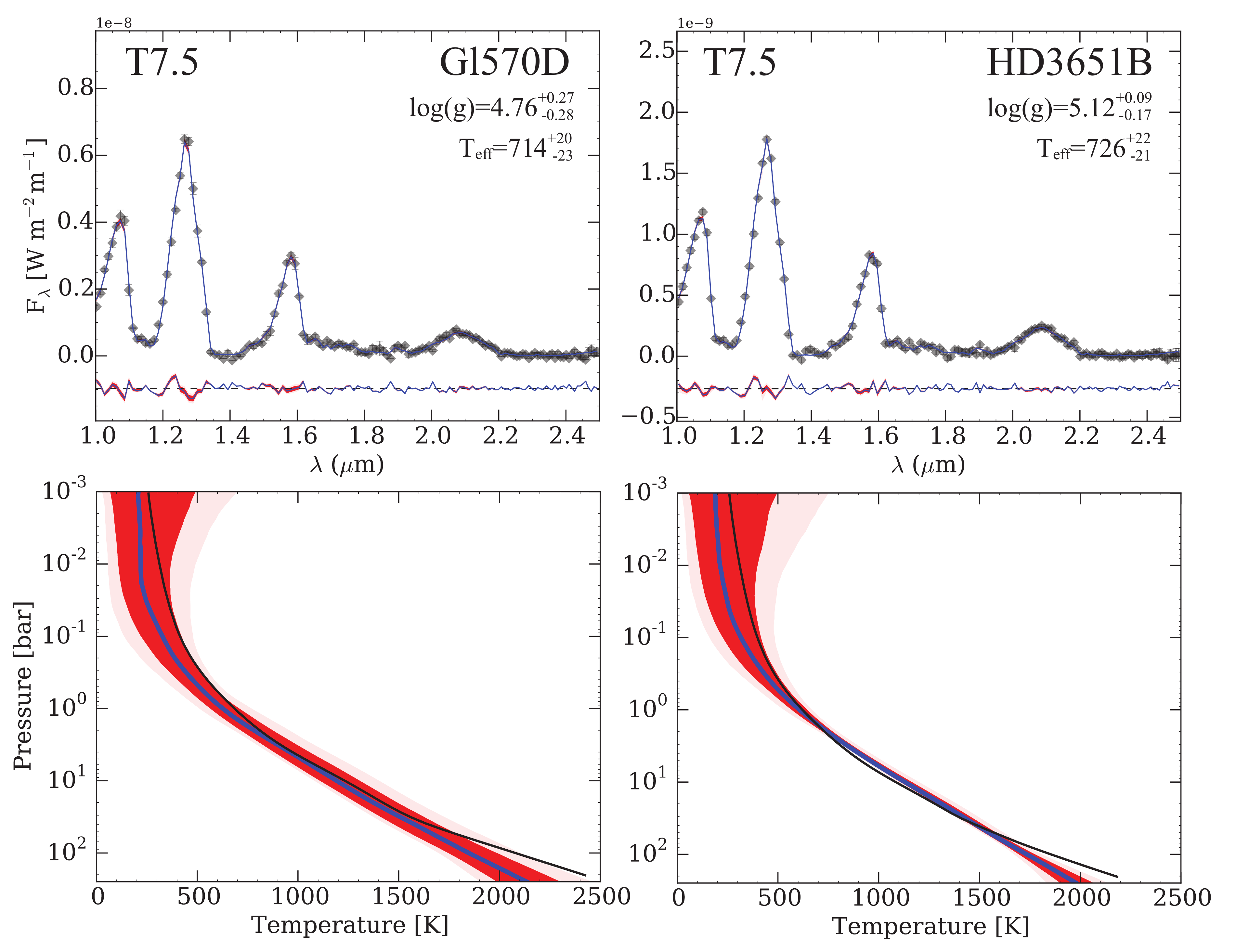}
%\end{center}
     \caption{ \label{fig:figure3}  Spectra (top row) and retrieved temperature profiles (bottom row).  For the two objects we show the H-band calibrated SpeX data as the diamonds with error bars, a summary of thousands of model spectra generated from the posterior and their residuals (median in blue, 1$\sigma$ spread in red), and their spectral type and bulk properties.  The bottom row summarizes thousands of temperature profiles drawn from the posteriors for each object (median in blue, 1$\sigma$ spread in red, 2$\sigma$ spread in pink).  The black temperature profile shown for each object is a representative self-consistent grid model (Marley et al. 2008) interpolated to the quoted $\log g$ and $T_{\rm eff}$ to demonstrate that our retrieved profiles are physical and are consistent with 1-D radiative convective equilibrium.       }
  \end{figure*}

\begin{figure*}
%\begin{center}
\includegraphics[width=1\textwidth, angle=0]{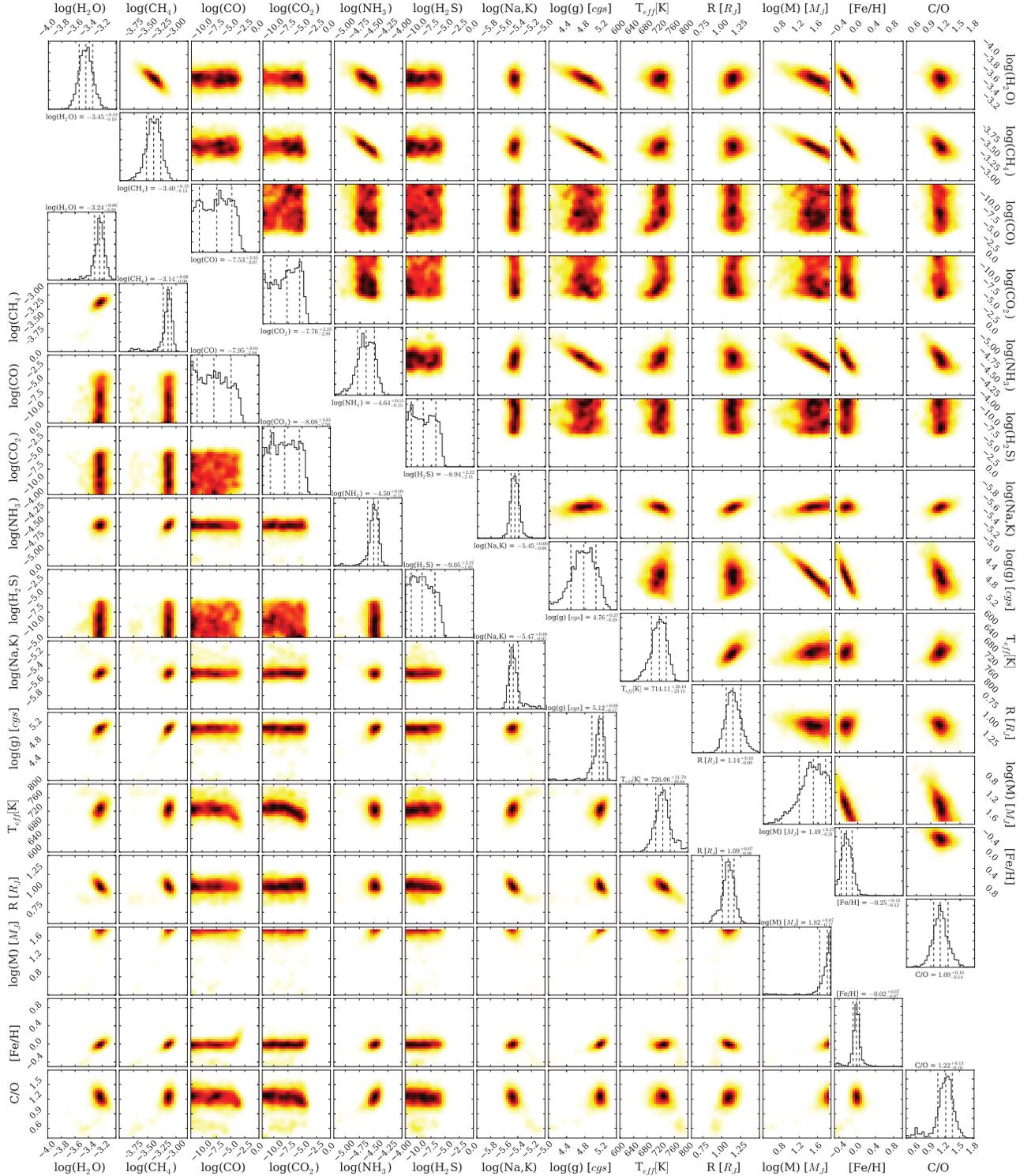}
%\end{center}
     \caption{ \label{fig:figure4} Summary of the posterior for the relevant parameters.  The stair-step plot on the top right is for Gl570D, and the bottom left for HD3651B.  These show the marginalized posteriors (1D histograms) along the diagonals and the parameter correlations (2D histograms).   The parameters for each object are on the same scale and so can be directly compared.  The dashed lines in the 1D histograms are the 16, 50, and 84 percentiles.  The width between the 16 and 84 percentiles represents the 68\% confidence interval.  For each parameter, the median and plus/minus 1$\sigma$ values are shown just above (HD3651B) and below (Gl570D) the histograms.         }
  \end{figure*}

The fiducial retrieval results are summarized in Figures \ref{fig:figure3} and \ref{fig:figure4}. The top row of Figure \ref{fig:figure3} shows an ensemble of thousands of fits derived from the posterior and their residuals.  We also summarize the retrieved $\log g$ and 
$T_{\rm eff}$.   The residuals appear to be minimal and random, with exception of the peaks near Y and J band.  The slight misfit at these wavelengths could potentially be due to uncertainties in the alkali (sodium, potassium) cross sections, which have the least certain cross-sections of the opacities that absorb at these wavelengths.  We note that these residuals are much smaller than can be obtained by typical grid model fits suggesting that the additional parameters we include in our model are required to produce these better fits.  

%%%%%%%%%%%%%figure3-TP/Spectra%%%%%%%%%%%%%%%%%%%%
%%%%%%%%%%%%figure3%%%%%%%%%%%%%%%%%%%%

The retrieved temperature profiles (bottom row, Figure \ref{fig:figure3}) for each object are summarized with a median, 1, and 2$\sigma$ credibility region from the ensemble of thousands of randomly drawn TP profiles from the posterior. The retrieved TP profiles appear to be consistent with physically based expectations.  In each panel we show for comparison a self-consistent grid model profile (black, from the grid models of Saumon \& Marley 2008) corresponding to the median retrieved effective temperature and gravity.  The self-consistent grid models assume cloud-free 1-dimensional radiative convective equilibrium with a solar composition atmosphere in thermochemical equilibrium.  The agreement is astounding and this is a point we would like to stress.  The self-consistent grid model profiles generally fall well within the 2$\sigma$ credibility region.  We have made no assumptions about the nature of the TP profiles  other than smoothness, the degree of which was allowed to vary.  We also tested different starting guesses (e.g., isothermal) and the results are no different.  This suggests that these are the actual, true, temperature-pressure profiles in these atmospheres and that the assumption of one-dimensional radiative-convection is sufficient.   However, there is some divergence at pressure levels less than about 1 bar and greater than a few 10s of bars. The retrieved profiles tend to become more isothermal near the top of the atmosphere than the radiative convective models, though this divergence is much smaller than the width of the confidence intervals and is therefore not significant (as constrained by SpeX data alone).  We also note that a recent paper by Tremblin et al. (2015) predicts that condensation-induced fingering convection can result in a cooler deep atmosphere than expected from standard dry convection assumptions, consistent with what we are finding but not conclusive.  Higher resolution data with more vertical resolution and altitude range or longer wavelength data will (and have, as in Line et al. 2014b) provide better constraints to the TP profile that will allow us to further test deviations from the standard radiative-convective assumptions.  In this investigation, we purposefully avoid combining different datasets in this investigation due to the introduction of additional systematics that come with multiple data sets.

Figure \ref{fig:figure4} summarizes the posterior for the physical atmospheric parameters. The effective temperature, radius, mass, metallicity ($[\rm Fe/H]$), and C/O ratio are all derived quantities and were not directly retrieved.  The effective temperature distribution is obtained by computing the bolometric flux (1-20 $\mu$m) over thousands of spectra generated from the posterior.  The radius is derived from the retrieved spectral scaling factor, $(R/D)^2$, given the distance. The distance and photometric uncertainties are formally propagated into the radius uncertainty via Monte Carlo error propagation.  The mass is derived from the retrieved gravity and radius.  We also reiterate that the radius, and hence the derived mass, are very sensitive to the accuracy (e.g., missing systematics not accounted for in the quoted photometric errors) of the photometry. We discuss the impact of the photometric calibration on the retrieved quantities further in $\S$\ref{sec:photometry}.  The metallicity is derived by summing up the molecular mixing ratios for each species weighted by the number of metal atoms divided by the abundance of hydrogen and then comparing that to the sum of solar metals relative to hydrogen.  The C to O ratio is computed by dividing the sum of the carbon bearing species by the oxygen bearing species appropriately weighted by the number of carbon/oxygen atoms in each species.

%%%%%%%%%%%%%figure4-Stair Pairs%%%%%%%%%%%%%%%%%%%%
%%%%%%%%%%%%figure4%%%%%%%%%%%%%%%%%%%%

The marginalized probabilities for each parameter (the histograms) are shown along the diagonals for each object. For both objects the H$_2$O, CH$_4$,  NH$_3$ and Na+K are constrained (68\% confidence) to better than 0.3 dex (or a factor of 2).  For comparison the best constraints we have on gases in exoplanet atmospheres are to within a factor of $\sim$ 10 (e.g., Kreidberg et al. 2014).    

Perhaps the most surprising finding is the tight constraint on ammonia for each object.   The robustness of the ammonia constraints in the near-IR are discussed in $\S$\ref{sec:ammonia}.  Additionally there are strong correlations of log$g$ with spectrally prominent absorbers, H$_2$O, CH$_4$, and NH$_3$ .  As expected for any atmosphere in hydrostatic equilibrium, there is a positive correlation between metallicity and gravity (this can also be seen in the metallicity vs. gravity panels in Figure \ref{fig:figure4}).  As gravity increases, the optical depth at a given layer in the atmosphere decreases ($\tau=\kappa P/g$, where $\kappa$ is the opacity), so the opacity must increase to maintain that same optical depth.   The strong absorbers, H$_2$O, CH$_4$, and NH$_3$, are also correlated with each other and with temperature (not shown).  These correlations are positive, which seems backwards for overlapping absorbers, but it is because as one absorber increases, the temperature first increases to maintain that same flux at the wavelength of that absorber, resulting in a higher flux at a different wavelength where another absorber is present.  This absorber must increase in abundance to suppress the flux at this different wavelength.   The other gases, CO, CO$_2$, and H$_2$S, are largely unconstrained; only upper limits can be obtained. This is mainly because of their relatively low thermochemical abundances (despite potential vertical mixing) and relatively weak bands in the near infrared.  We would expect CO and CH$_4$ to flip roles for objects hotter than $\sim$1100 K.

In order to check whether or not the retrieved abundances are chemically realistic, we compare them to thermochemical equilibrium models. To do this, we use the NASA Chemical Equilibrium with Applications model (CEA, Gordon \& McBride 1996) recently used for exoplanet atmosphere studies (Line et al. 2010; Moses et al. 2011; Line et al. 2011).  CEA only requires the local temperature, pressure, and elemental abundances at given model layer in order to compute the equilibrium abundances.  Equilibrium condensation chemistry (no rainout) is included but the code has a difficult time with enstatite (MgSiO$_3$) condensation. For this reason we do not include magnesium or silicon species, but account for the depletion of oxygen due to presumed cloud formation in the deep atmosphere by removing 3.28 oxygen atoms for every silicon atom (Burrows \& Sharp 1999).  We have not accounted for perturbations to equilibrium chemistry such as horizontal or vertical mixing.  

\begin{figure*}
%\begin{center}
\includegraphics[width=1\textwidth, angle=0]{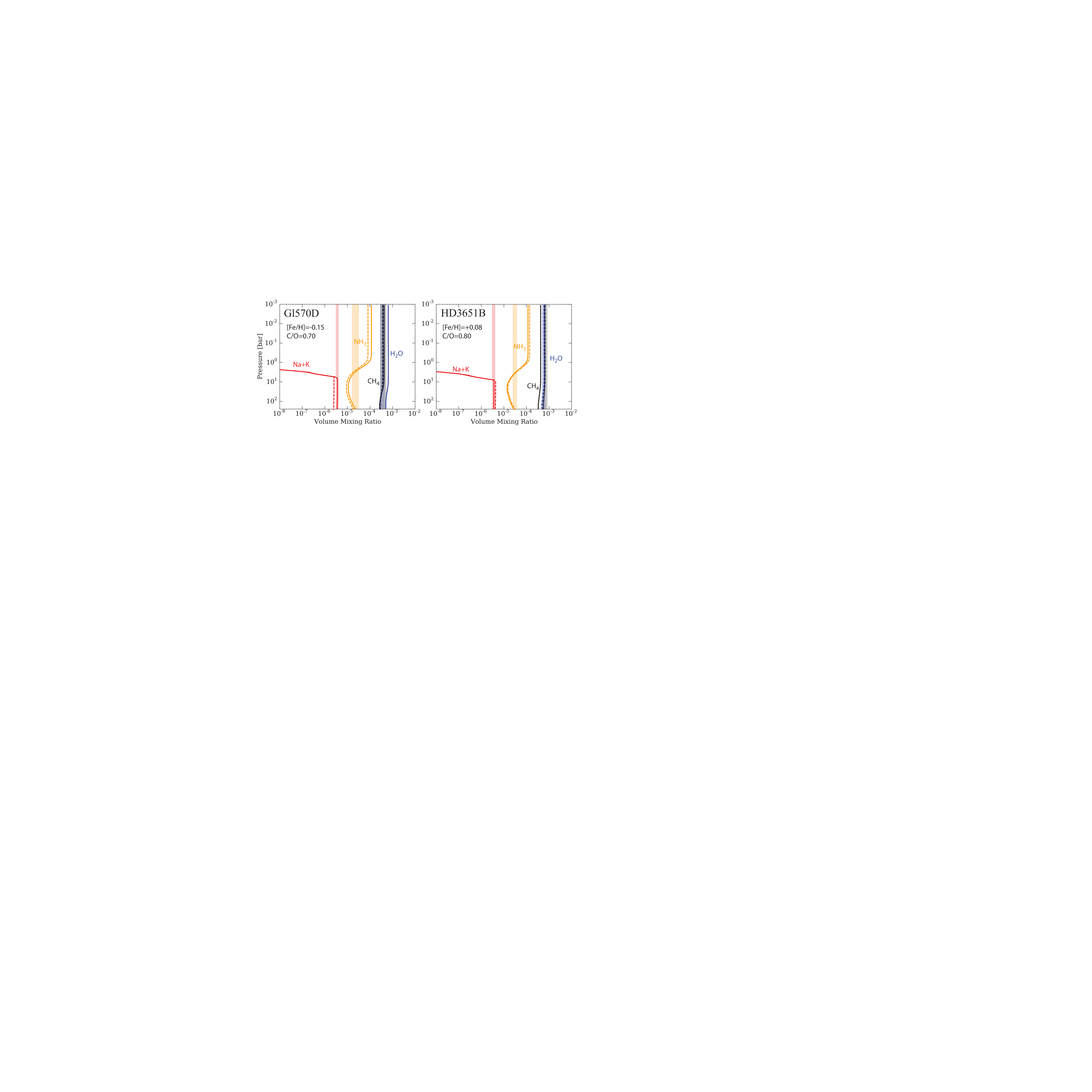}
%\end{center}
     \caption{ \label{fig:figure5}  Comparison of the retrieved values (shaded boxes) of the well constrained molecules with their expected thermochemical equilibrium abundances along the median temperature profile. The solid curves are the thermocemical equilibrium abundances for solar composition while the dashed curves are the thermochemical equilibrium abundances for the specified C/O and metallicity.  This shows that the retrieved abundances are thermochemically consistent.       }
  \end{figure*}

Figure \ref{fig:figure5} compares the retrieved results for the well constrained species to thermochemical equilibrium abundances along the median temperature profile.  In principle there will be a spread, albeit minor, due to uncertainties in the temperature profile, but since the goal is to just check for realism in the abundances we ignore such a spread.  For each object we show two cases of equilibrium abundances. The first is for solar elemental composition (solid lines) and the second is a by-hand fit of the $intrinsic$ metallicity and C to O ratio to the retrieved quantities.  We stress that the $intrinsic$ metallicity and C/O are different from the $atmospheric$ metallicity and C/O. Condensate processes can deplete oxygen or other species in the atmosphere resulting in $atmospheric$ metallicity and C/O's that differ from the $intrinsic$ or bulk values.   

We find for both Gl570D and HD3651B that the assumption of $intrinsic$ solar elemental abundances overestimates the retrieved water and methane abundances but appears to do a good job for the alkali and ammonia mixing ratios.  However, by hand-tuning the bulk metallicity and C/O in the thermochemical model we can better match the retrieved water and methane abundances.  This is not a rigorous ``fit" to the chemistry by any means, but simply an attempt to show that we have retrieved chemically plausible molecular abundances.  A perhaps more rigorous approach for obtaining the allowed ranges of C to O ratios and metallicities would be to perform a ``retrieval on the retrieval" where by the chemical model would fit the retrieved molecular abundances within a Bayesian framework. This is currently beyond the scope if this work.  We also note that the ammonia thermochemical profiles agree well with our column averaged uniform-with-altitude retrieved values. Saumon et al. (2006) suggest that quenching of ammonia due to vertical mixing occurs in the deep atmosphere near a temperature of $\sim$2200 K. Such temperatures occur at the deepest pressures (several hundred bars) on our retrieved profiles. If ammonia indeed quenches at these deep levels then the ammonia profile would be nearly constant with altitude well within our retrieved range for both objects (Figure \ref{fig:figure5}).

Another surprising find is that the retrieved water/methane abundance is lower (water/methane $\sim$ 0.8-0.9) than what one may expect for a solar composition atmosphere. Typically water is more abundant than methane by a factor of $\sim$1.5 at solar composition (see Saumon et al. 2006).  This suggests a super-solar (greater than 0.5) atmospheric carbon-to-oxygen ratio. Accounting for the draw down of O due to silicate condensation, the methane/water abundance shown in Saumon et al. (2006) for Gl570D suggests an $atmospheric$ C/O of 0.63.  Our retrieved $atmospheric$ carbon-to-oxygen ratios for both objects are higher than one.   The inferred, by-hand $intrinsic$ C to O ratios are less than unity but are still slightly higher than solar (see \S \ref{sec:benchmark} for a comparison to the host star values). We note that, in our previous study, Line et al. (2014b), we found a fairly low ($\sim$0.2) carbon-to-oxygen ratio for Gl570D. These differences are likely due to the inclusion of the Akari and Spitzer Infrared Spectrometer data and the treatment of the systematics between them.  Our current investigation is more straightforward as we only focus on the SpeX data set, and thus do not have to worry about potential biases due to differing unaccounted for systematics amongst different datasets. In $\S$\ref{sec:benchmark} we show that our thermochemically self-consistent $intrinsic$ carbon-to-oxygen ratios for the brown dwarf's are consistent with the those derived from the stellar primaries.

%%%%%%%%%%%%%figure5-Chemistry%%%%%%%%%%%%%%%%%%%%

\subsection{Ammonia in the Near-IR}\label{sec:ammonia}
One of the more remarkable findings in this investigation is the strong evidence for the presence of ammonia in {\it low resolution near infrared spectra}.  Ammonia at longer wavelengths in T-dwarfs is not new. Cushing et al. (2008) and Saumon et al. (2006) convincingly demonstrated the presence of ammonia in Gl570D using the strong 9.6 $\mu$m band in the Spitzer IRS data.  However, spectroscopic features of ammonia are not expected to present themselves below 2.5 microns until the Y-dwarfs (Kirpatrick et al. 2005) unless observed at high resolution (Canty et al. 2015).   Therefore, we were surprised to find how strongly ammonia can be constrained (e.g., an actual bounded limit as opposed to an upper limit only) with $low$-resolution near infrared data in both objects despite the lack of 
obvious spectral features in this wavelength range.   How are we to believe that our constraint is real?  We show three lines of evidence supporting our strong ammonia constraint.  

One line of evidence comes from a standard Bayesian hypothesis testing procedure. Such procedures  determine whether or not a parameter within nested models is justified given the data (e.g., Trotta et al. 2008). A commonly used approach is to compare the Bayesian information criterion (BIC) between a model with and without a particular parameter.  Parameters that provide better fits to the data and produce a delta-chi-square that is greater than the penalty of adding that additional parameter are justified. However, the BIC is a truncated Laplace approximation to the full Bayesian evidence, or marginal likelihood (Kass \& Raferty 1995).   We do not use the BIC here, rather we compute the full Bayesian evidence.  This is done by numerically integrating over the entire posterior using the approach described in Weinberg et al. (2012) (see also Swain, Line \& Deroo 2014 for an application to exoplanet spectra).  By computing the evidence of the full model that contains all parameters to the one that removes ammonia, we can obtain a Bayes factor. Bayes factors greater than one suggest that the model containing the parameter in question is favored, while Bayes factors less than one suggest otherwise.  A Bayes factor can then be converted into a confidence of detection (Trotta 2008).  Table \ref{tab:table3} shows the Bayes factors and the corresponding detection significances for three different nested models each removing only one gas (NH$_3$, H$_2$S, or H$_2$O) from the full model in Table \ref{tab:table1}. 

We show the detection significances for H$_2$O as an example of a gas that is visibly obvious in the near infrared and well constrained (Figure \ref{fig:figure4}), and hence we would expect an extremely high detection significance.  We detect water at at an extremely high degree of confidence,  $>$17 $\sigma$, in both objects.  H$_2$S is an example of a poorly constrained species (Figure \ref{fig:figure4}), and hence would expect, and indeed do find, a low detection significance below 2$\sigma$ .  In fact for, HD3651B the Bayes factor is less than one ($ln$(Bayes factor) $<$ 0) or evidence against H$_2$S.  Since we don't visibly see any obvious spectroscopic features due to NH$_3$, but do indeed obtain a strong constraint (the marginalized posterior is bounded on both sides as opposed to an upper limit like H$_2$S), we may expect the detection significance to fall in between the two aforementioned extremes.  This is what we do indeed find, a  $>$6$\sigma$ detection of NH$_3$ in both objects, which is considered strong.  We should note that Bayes factors can be sensitive to the prior ranges. We found that in our case, the Bayes factor calculation is insensitive to our prior ranges.

\begin{figure*}
%\begin{center}
\includegraphics[width=1\textwidth, angle=0]{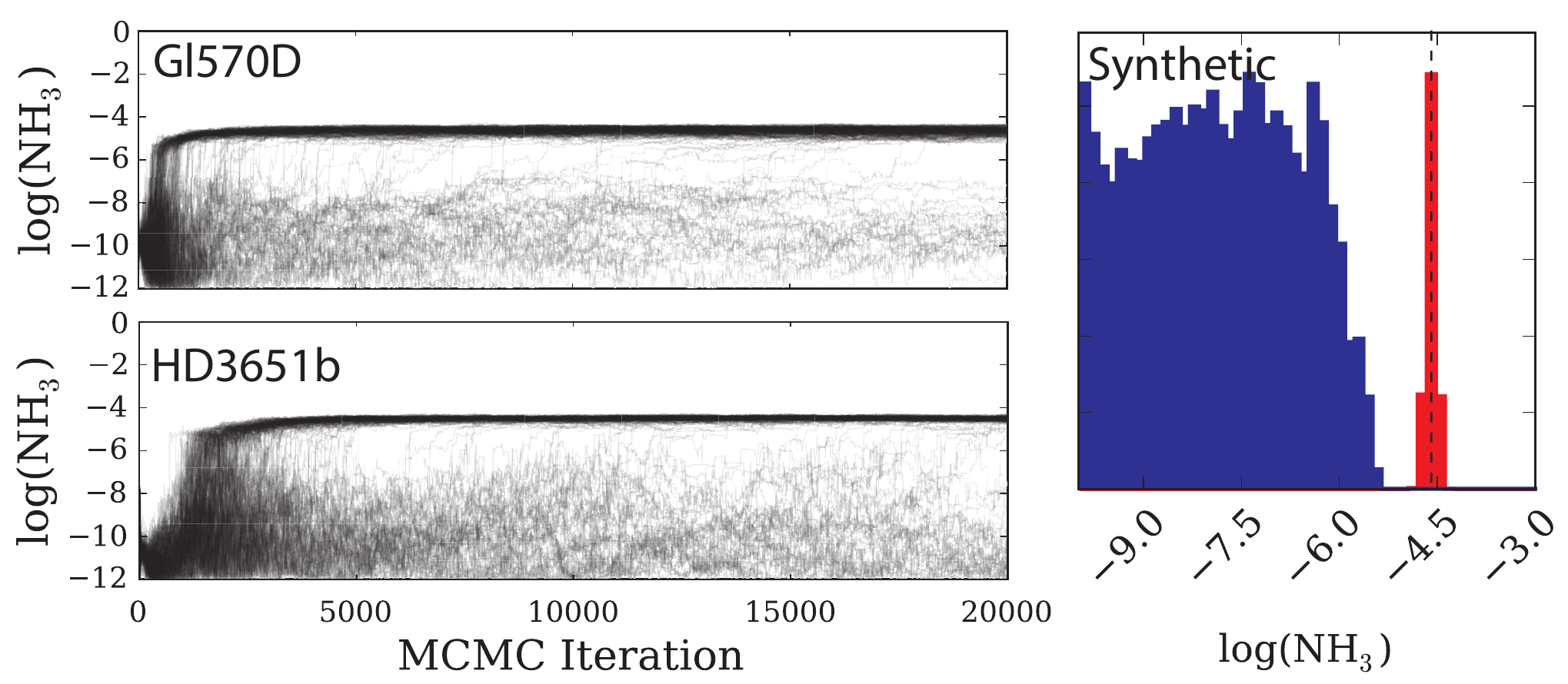}
%\end{center}
     \caption{ \label{fig:figure6}  Evidence for ammonia in the near infrared.  The left two panels show show the evolution of the MCMC chains initialized at a non-detectable value.   The chains for each object readily converge towards a well constrained solution about the quoted retrieved values.  The outlier chains account for less than 10$\%$ of the total probability.  The right panel shows the retrieved probability distribution of ammonia for two $synthetic$ brown dwarf spectra.  The first synthetic spectrum was generated with a mixing ratio of $10^{-10}$. The retrieved probability (blue) shows only an upper limit--this is consistent with a non-detection--as we would expect.  The second synthetic spectrum was generated with an ammonia mixing ratio similar to the retrieved values for Gl570D and HD3651B (vertical dashed line).   The retrieved probability distribution (red) is bounded on both sides about the truth suggesting a strong constraint.    }
  \end{figure*}

We also show two additional, more straight forward, lines of evidence in Figure \ref{fig:figure6}. For the first test we re-ran the retrieval but initialized the MCMC with a a non-detectable ammonia abundance far from the retrieved value.  If the retrieved value we obtain is true, then we would expect the ensemble of Markov chains to converge towards the true value regardless of the starting point--given a long enough run time.  That is indeed what we find.  The left two panels in Figure \ref{fig:figure6} show the evolution of the Markov chains.  They readily rebound from the poor initial starting point (essentially no ammonia) and converge to a nice tightly packed bundle within the target distribution.  Less than 10$\%$ of the chains remain outliers. Had we run for even longer (weeks perhaps) these chains would likely have fallen in line with the rest and too converge to the true answer.  

For the final test for ammonia we create two synthetic brown dwarf spectra for which we know the true TP profile and composition.  We choose parameter values and data properties similar to Gl570D.  We create two synthetic spectra: one with ammonia and one without. We then apply the retrieval to these two spectra. What is shown in the right panel Figure \ref{fig:figure6} are the retrieved ammonia distributions for these two scenarios.  The blue histogram is the retrieved ammonia distribution for the synthetic spectra generated $without$ ammonia.  In this case only an upper limit of ammonia can be obtained. This means that there are no spectral features present in the synthetic spectrum to suggest the existence of ammonia. When the ammonia abundance creeps up to a high enough values (in this case $\sim$ 1ppm) it begins to present itself in the spectrum in an undesirable fashion.  The red histogram is the retrieved ammonia distribution for the synthetic spectra generated $with$ ammonia. This histogram is nicely bounded on both sides suggesting that the spectral features due to ammonia are enough to place both a lower and upper bound on the retrieved abundance. These three lines of evidence strongly suggest the presence of ammonia in the the low-resolution near infrared spectra of these two T-dwarfs. 
 
%%%%%%%%%%%%%figure6-Ammonia Evidence%%%%%%%%%%%%%%%%%%%%
%%%%%%%%%%%%figure6%%%%%%%%%%%%%%%%%%%%

%%%%%%%%%%%%%%%%%%%%%%%%%%Table3-NH3 Evidence%%%%%%%%%%%%%%%%%%%%
\begin{table}
\centering
\caption{\label{tab:table3} Bayesian nested model comparison supporting the presence of ammonia in the near IR.  Three model scenarios are shown, one that includes all parameters except NH$_3$ ( NH$_3$), one that includes all parameters except H$_2$S (H$_2S$), and one that includes all parameters but H$_2$O (H$_2$O).  The Bayes factors are computed relative to the full model which includes all of the parameters shown in Table \ref{tab:table1} . }
%\resizebox{\textwidth}{!}{%
\begin{tabular}{cccccc}
%\footnote{\tiny{aa}}
\hline
\hline
\cline{1-2}
                &             ~~Gl570D    &        &           ~~HD3651B     & \\
Scenario  & ln$B$ & Det. Sig. & ln$B$ & Det. Sig.     \\
                &             &  ($\sigma$)               &        & ($\sigma$) \\
\hline
NH$_3$ 	& 20.5	& 6.7		&	20.7	&  6.8\\
H$_2$S 	& 0.7	         & 1.8	&	-1.7		&  -\\
H$_2$O 	& 166.1	& 18.4	&	153.4		&  17.7\\

\hline
\end{tabular}
\end{table}
%%%%%%%%%%%%%%%%%%%%%%%%%%Table3%%%%%%%%%%%%%%%%%%%%

\subsection{Verifying Cloud Free}
T-dwarfs are typically assumed to be cloud free given their blue colors (Burrows et al. 1997; Allard et al. 2001) and the success of cloud-free grid models to reasonably explain their spectra (e.g., Stephens et al. 2009), though some of the redder objects are better matched with models that include sulfide-like clouds (Morley et al. 2012).  Since we assumed cloud free atmospheres for our retrieval conclusions, we need to make sure that this is indeed the case.  We are not interested in the cloud properties themselves, rather the impact that some unaccounted for gray absorber may have on the spectra.  Therefore we model clouds rather simplistically as a gray absorber with opacity $\kappa_{c}$ between two defined pressure levels, $P_{c,\text{bottom}}$ and $P_{c,\text{top}}$.  The cloud optical depth is 
\begin{equation}
\tau_{c}=\kappa_{c}\frac{P_{c,\text{bottom}}-P_{c,\text{top}}}{g}
\end{equation}
We assume $P_{c,\text{bottom}}-P_{c,\text{top}}$ spans one scale height so that we only need to retrieve the cloud base location ($P_{c,\text{bottom}}$) and $\kappa_{c}$.  The gray approximation can be readily justified.  From grid model investigations with clouds (e.g., Morley et al. 2012) high sedimentation values (Ackerman \& Marley 2001) are required to best match the spectra.  High sedimentation values generally result in a wider range of particle sizes thus washing out Mie scattering features resulting in gray, or at least, nearly gray absorption.  

\begin{figure*}
%\begin{center}
\includegraphics[width=1\textwidth, angle=0]{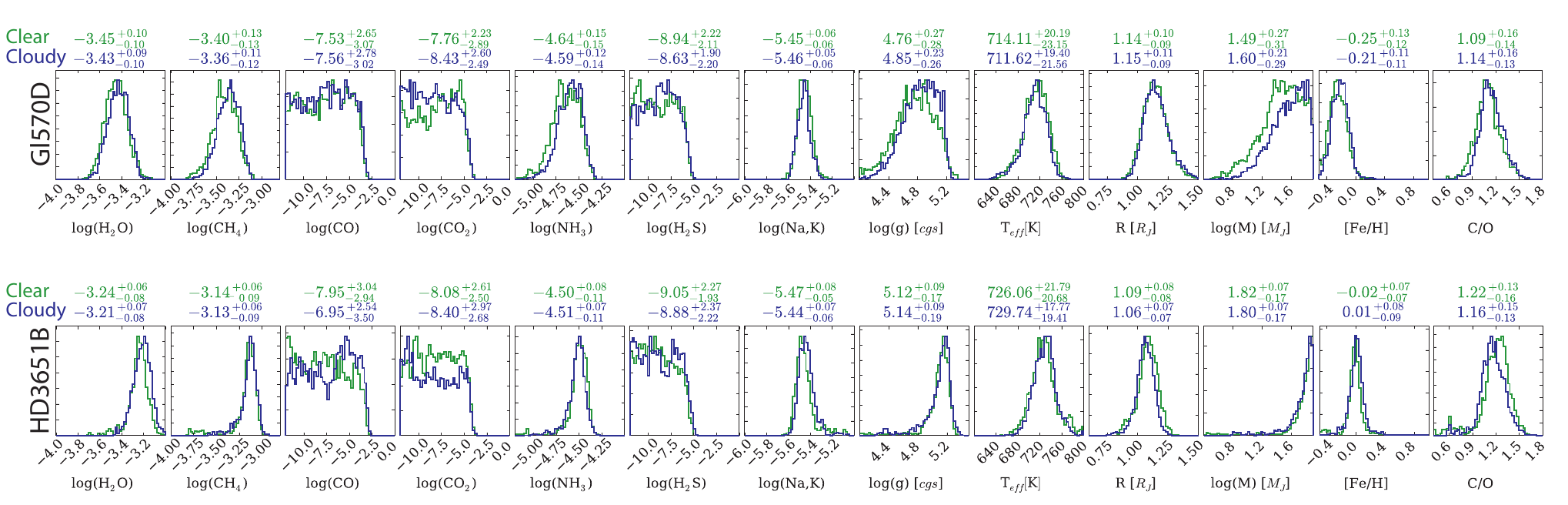}
%\end{center}
     \caption{ \label{fig:figure7} Impact of a gray cloud on the retrieved quantities for Gl570D (top row) and HD3651B (bottom row).   In green we show the retrieved marginalized posterior distributions for the cloud free nominal model as in Figure \ref{fig:figure4} and in blue for the cloudy model. The median and 1$\sigma$ confidence interval for each scenario are shown above each histogram. The shift in the medians of all parameters remain less than the one sigma uncertainty suggesting that the inclusion of a gray cloud has a minimal impact.    }
  \end{figure*}

In order to determine whether or not clouds are present, we undergo the same Bayesian hypothesis testing procedure described in $\S$\ref{sec:ammonia}. This time the full model includes all of the parameters in Table \ref{tab:table1}, plus the two cloud parameters. We compare the evidence of original cloud free model to this new full model.  Table \ref{tab:table4} shows the Bayes factor and detection significance for the cloud.  We find that for Gl570D the detection significance is below 2 $\sigma$ suggesting a weak detection of clouds.  HD3651B presents a slightly higher, or weak to moderate detection of a cloud.   Figure \ref{fig:figure7} shows the parameter distributions for both the cloudy and clear atmospheres; including the clouds has an insignificant impact on all of the retrieved parameter values (e.g., the change in the median value is less than the typical width of the distributions).   Therefore, we are justified in assuming cloud free atmospheres for these two objects.

%%%%%%%%%%%%%%%%%%%%%%%%%%Table4-Cloud Evidence%%%%%%%%%%%%%%%%%%%%
\begin{table}[h]
\centering
\caption{\label{tab:table4} Bayesian nested model comparison demonstrating lack of evidence for clouds.  The Bayes factors are computed relative to the model which includes all of the parameters shown in Table \ref{tab:table1} $and$ the two additional cloud parameters (see text).  }
%\resizebox{\textwidth}{!}{%
\begin{tabular}{cccccc}
%\footnote{\tiny{aa}}
\hline
\hline
\cline{1-2}
                &             Gl570D    &        &           HD3651B     & \\
Scenario  & ln$B$ & Det. Sig. & ln$B$ & Det. Sig.     \\
                &             &  ($\sigma$)               &        & ($\sigma$) \\
\hline
Cloud 	& 0.76	& 1.87		&	1.65	& 2.38\\
\hline
\end{tabular}
\end{table}
%%%%%%%%%%%%%%%%%%%%%%%%%%Table4%%%%%%%%%%%%%%%%%%%%

%%%%%%%%%%%%%figure7-Cloud vs. No cloud%%%%%%%%%%%%%%%%%%%%
%%%%%%%%%%%%figure7%%%%%%%%%%%%%%%%%%%%

\subsection{Impact of Photometry}\label{sec:photometry}
\begin{figure*}
%\begin{center}
\includegraphics[width=1\textwidth, angle=0]{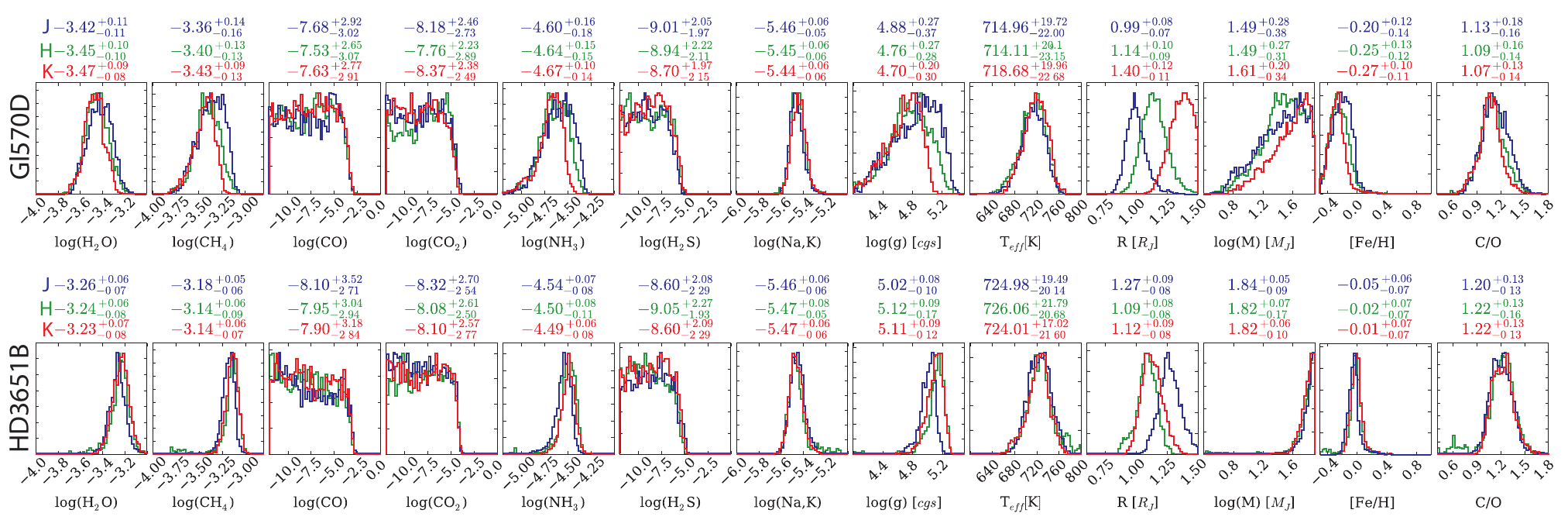}
%\end{center}
     \caption{ \label{fig:figure8} Impact of choice of photometry used to calibrate the normalized Spex spectra.  For each object (Gl570D-top row, HD3651B-bottom row) we show the marginalized posteriors resulting from the calibrated spectra from each of the three photometric band (blue for J-band, green for H-band (nominal case), and red for K-band). We also show the median and 1$\sigma$ confidence interval for each parameter for each photometric case.      }
  \end{figure*}

The flux calibrated spectra depend on the choice of photometry used as demonstrated in Figure \ref{fig:figure1}.  Here, we explore the impact on the retrieved quantities of the choice in photometry.  For each object we calibrate the spectrum with either the J-band, H-band, or K-band 2MASS photometry, as shown in Figure \ref{fig:figure1}.  As mentioned earlier, Stephens \& Leggett (2004) provide correction factors for the 2MASS photometry. We do not apply those correction factors here, rather the goal is to determine what effects ``bad" photometry may have on the retrieved quantities.  We then execute the retrieval on each of those three calibrated spectra.  Figure \ref{fig:figure8} shows the resulting retrieved quantities for each of the photometric calibration scenarios.  The impact is minimal for most quantities (e.g., the shift in the median is well within the 1$\sigma$ uncertainties) with the exception of the photometric radius. This is unsurprising as the overall scaling to the spectrum depends on $(R/D)^{2}$. Shifts in this scaling due to photometry will result in changes in the derived radius. We also find small ($\sim$ 1$\sigma$) shifts in the retrieved gravity due to the prior upper limit on the mass (masses cannot exceed 80 M$_{J}$). This is because the mass depends on both the radius and gravity therefore the small shifts in the radius propagate through the mass upper limit to the gravity. If the radius increases due to a change in photometry, then the gravity has to decrease. These shifts in derived radius and gravity stress the importance of precision photometry on these objects.

%%%%%%%%%%%%%figure8-Photometry%%%%%%%%%%%%%%%%%%%%
%%%%%%%%%%%%figure8%%%%%%%%%%%%%%%%%%%%

\section{Validating Retrieval with Benchmarks}\label{sec:benchmark}
Both Gl570D and HD3651B happen to orbit stars with known properties. This makes them powerful benchmark systems for which we can test the validity of our retrieval approach.  

The basic properties we use to evaluate the benchmark systems are their evolutionary derived ages, metallicities, and the carbon-to-oxygen ratios. A fundamental assumption with benchmark systems is that the primary and companion both formed out of the same nebular material at the same time and should each individually indicate the same elemental abundances and age.  This is unlike planetary systems in which we believe that planet formation processes within the protoplanetary disks can alter the planetary atmosphere abundances relative to their host star (Oberg et al. 2011; Fortney et al. 2013). 

 A summary of the relevant stellar primary and retrieved brown dwarf properties are shown in Table \ref{tab:table5}.  Since stellar photospheres are generally assumed to be well mixed and free from condensates, their photospheric abundances are representative of their $intrinsic$ values, however for the brown dwarfs we retrieve the $atmospheric$ quantities rather than the $intrinsic$ quantities, which can be different due to the aforementioned reasons.  Therefore, in Table \ref{tab:table5} we show both the $atmospheric$ elemental quantities and the thermochemically self-consistent $intrinsic$ elemental quantities derived from the hand tuned fits (Figure \ref{fig:figure5} ) of the chemical model to the retrieved abundances.  Typically the atmospheric oxygen is depleted by 20-30$\%$ relative to the $intrinsic$ due to the sequestration of oxygen in condensates (depending on the $intrinsic$ metallicity and C/O).  This results in a higher $atmospheric$ carbon to oxygen ratio and an overall lower atmospheric metallicity, since oxygen is the dominant metal atom. In Table \ref{tab:table5} we list the $inferred~intrinsic$ metallicity and C/O based on the atmospheric measurements and the estimated depletion of O and silicates due to condensation. These are the nominal values against which we compare to the stellar abundances.    We find that the metallicities in both objects are somewhat lower than what is measured in the primaries but are still consistent.   From the C/O measurements in both the brown dwarf ($\S$\ref{sec:appendix}) and host star, we have an additional benchmark constraint.  We find that for the Gl570 system, the 1$\sigma$ inferred intrinsic C/O range falls entirely within the stellar primary C/O values.  We consider this an excellent agreement.  For the HD3651 system, the inferred intrinsic C/O is higher than the stellar primary by $\sim$30\%, however their medians are consistent at the 2$\sigma$ level.   This represents, for the first time, a comparison of a stellar and companion brown dwarf carbon-to-oxygen ratios.   This suggests that we are in fact retrieving the proper molecular abundances in the brown dwarf atmospheres. 
 
\begin{figure}
%\begin{center}
\includegraphics[width=0.5\textwidth, angle=0]{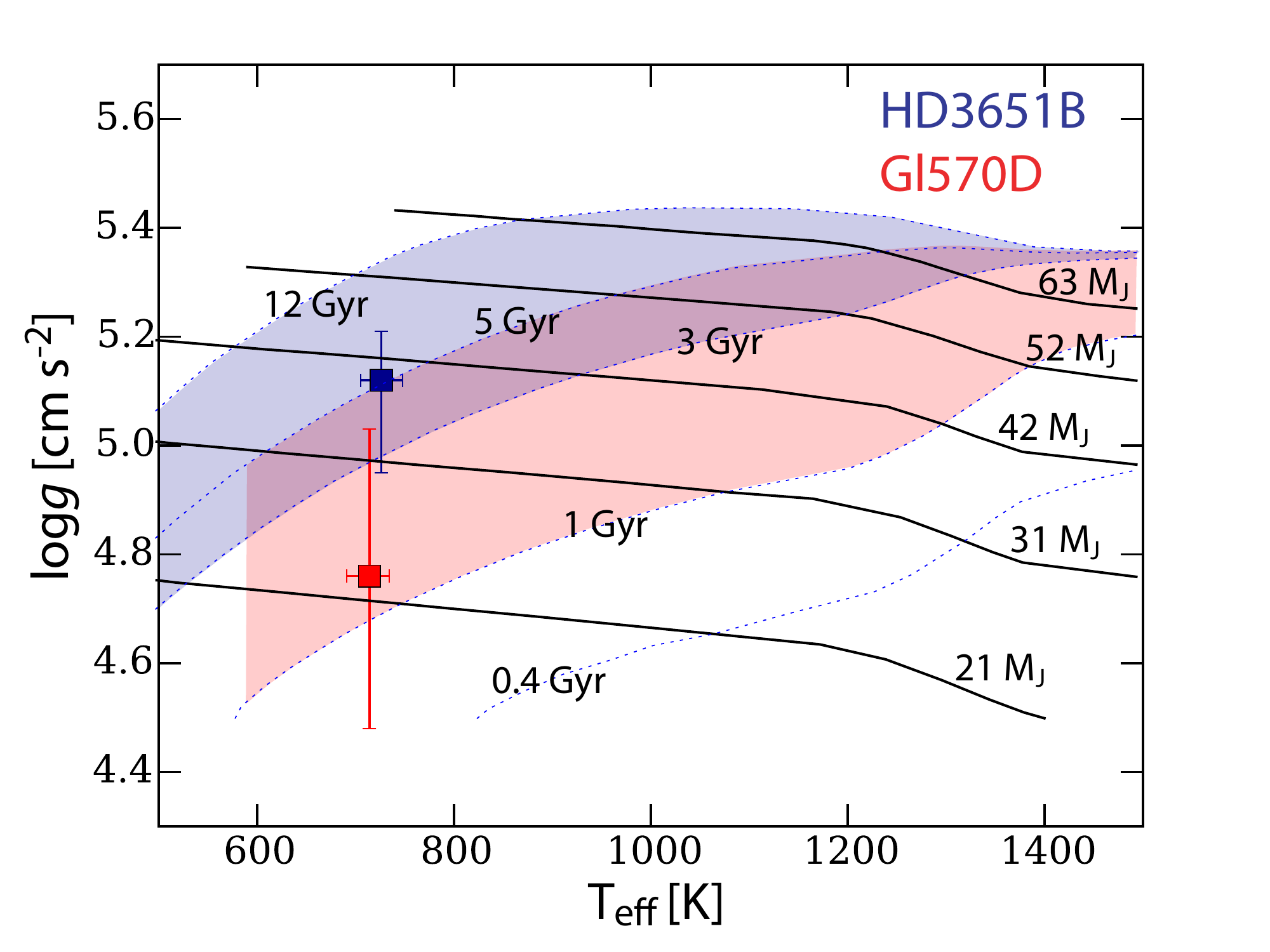}
%\end{center}
     \caption{ \label{fig:figure10} Comparison of our retrieved gravity and effective temperature to the evolution tracks of Saumon \& Marley (2008).    Our retrieved values are the red (Gl570D) and blue (HD3651B) boxes with error bars.  The dotted lines are the log$g$-T$_{eff}$ isochrones.  The red and blue shaded regions are the range of estimated ages of the Gl570 and HD3651B systems, respectively.  The inferred evolutionary ages are consistent with the estimated system ages.  }
  \end{figure}

 Finally, we compare the evolution-derived age to the estimated system age.  Figure \ref{fig:figure10} shows the Saumon \& Marley (2008) isochrones in log $g$-effective temperature space.  The shaded regions represented the estimated system ages as described extensively in  Liu, Leggett, \&  Chiu (2007).  Our retrieved gravity and effective temperature are consistent with the presumed system ages.  We also obtain photometric masses from the retrieved gravity and photometric radii.  We find that the 1$\sigma$ range in photometric masses for Gl570D (15-58 M$_J$)  is consistent with the evolution model masses (Figure \ref{fig:figure10}); however for HD3651B we find somewhat higher photometric masses (45-78 M$_J$) than anticipated from the evolution models. It is unclear why this may be.

In summary, we find ages derived from our retrieved gravity and effective temperatures are consistent with the measured system age and that the retrieved metallicities and C to O ratios, after taking into account the loss of oxygen due to condensates,  are in good agreement with the primaries.
 
%%%%%%%%%%%%%%%%%%%%%%%%%%Table5-Benchmark Properties%%%%%%%%%%%%%%%%%%%%
\begin{table}
\centering
\caption{\label{tab:table5} Benchmark system properties (parameters from Liu, Leggett, \&  Chiu (2007) unless otherwise noted).  The C to O ratio for Gl570A is from our stellar abundance analysis described in the Appendix. Quantities labeled with ``retrieved" are the retrieved values from this study. }
%\resizebox{\textwidth}{!}{%
\begin{tabular}{lcc}
%\footnote{\tiny{aa}}
\hline
\hline
\cline{1-2}
Property	& 	HD3651B	&	Gl570D\\
\hline
Spectral Type................................	& T7.5	& T7.5\\
Host star spectral type..................	& 	K0V	&	K4V\\
Distance (pc)\footnote{van Leeuwen et al. 2007}...............................	&	11.06$\pm$0.03	&	5.84$\pm$0.03\\
Estimated age (Gyr)......................&3 - 12	& 1 - 5	\\
Host star [Fe/H]............................& 0.11 - 0.25\footnote{Ramierez et al. (2013)} & -0.22 - 0.12\footnote{derived from our stellar abundance analysis described in the Appendix} \\
Retrieved Atmospheric [Fe/H]........&  -0.09 - 0.05 & -0.37 - -0.12   \\
Chemically Derived Bulk [Fe/H]\footnote{the chemically derived bulk quantities are the ``by hand" thermochemical model fits to the retrieved molecular abundances}........& +0.08 & -0.15  \\
Inferred Intrinsic [Fe/H]\footnote{Assumes an atmospheric metal depletion due to loss of O and silicates of 20\% based on the chemical models}........& -0.01 - 0.13 & -0.29 - -0.04  \\
Host star C/O\footnote{derived from our stellar abundance analysis described in the Appendix}............................& 0.51 - 0.73  & 0.65 - 0.97 \\
Retrieved Atmospheric C/O.........&   1.06 - 1.35 & 0.95 - 1.25   \\
Chemically Derived Bulk C/O.........& 0.80 & 0.70  \\
Inferred Intrinsic C/O\footnote{assuming an O depletion from silicates of 28\% for HD3651B and 26\% for Gl570D based on the chemical model results }.........&   0.76 - 0.97 & 0.70 - 0.93   \\
Retrieved T$_{eff}$ [K].......................& 726$^{+22}_{-21}$    &  714$^{+20}_{-23}$ \\
Retrieved log$g$ [cm s$^{-2}$].................& 5.12 $^{+0.09}_{-0.17}$    &  4.76$^{+0.27}_{-0.28}$ \\
Retrieved Mass [M$_{J}$].....................& 66 $^{+12}_{-21}$   &31 $^{+27}_{-16}$ \\
Retrieved Radius [R$_{J}$].....................& 1.09 $^{+0.08}_{-0.08}$   &1.14 $^{+0.10}_{-0.09}$ \\
\hline
\end{tabular}
\end{table}

%%%%%%%%%%%%figure9%%%%%%%%%%%%%%%%%%%%

\section{Summary \& Conclusions}\label{sec:Conclusions}
We have established a new minimal-assumption retrieval approach for brown dwarf atmospheres that significantly advances the work of our previous retrieval study (Line et al. 2014b).  The new retrieval approach relies upon a more robust Bayesian estimator, forward model, temperature profile parameterization, a single continuous spectrum, and treatment of unknown systematic uncertainties permitting generous uncertainty estimates. Differences in our results compared with Line et al. (2014b) are likely due to these major changes.    From our new approach applied to two benchmark late T-dwarfs, Gl570D and HD3651B, we determined the allowed range of the thermal structures, molecular abundances, gravities, and radius-to-distance scalings directly from the data.  We found that this parameter set provides very good fits to the data in the form of minimal, nearly random, residuals.  We validated the chemical plausibility of the retrieved molecular abundances using a a well vetted thermochemical equilibrium model.  

Perhaps the most significant highlight of our work is the robust detection of ammonia in the low resolution near-infrared spectra in these late T-dwarfs. We presented three lines of evidence to support this claim.  Furthermore, we showed that clouds play a minimal role in sculpting the spectra of these two objects and their inclusion had little to no influence on the other parameters.  We also suggested that large systematic uncertainties in photometry can result in biased estimates of the photometric radii. 

 An additional highlight, is for the first time, using the carbon-to-oxygen ratio of the host-companion as an extra dimension in establishing benchmark systems.  We found a remarkable agreement of the carbon-to-oxygen ratios derived from a stellar abundance analysis for Gl570A and our retrieval analysis on Gl570D, and a consistent agreement within the HD3651 system (considering the spread in literature C/O values).   This is quite the accomplishment as two completely separate techniques on two different objects for which we would expect similar abundances, are in good agreement. This further bolsters the suggestion of Fortney (2012) to explore the role of C/O in T-dwarf atmospheres and that the C to O ratio should be considered as an additional dimension when interpreting brown dwarf spectra.  Finally, we found that the ages derived from the evolution models and our retrieved gravity and effective temperatures are consistent with the estimated system ages. It would be interesting in future investigations to identify systems for which the companion-primary C/O and metallicities differ by a significant amount. This could point to new physics and/or chemistry operating in substellar atmospheres.

This investigation further establishes the power of our novel retrieval approach in understanding the atmospheric and bulk properties of brown dwarfs.  In a future study we plan to apply this technique to a wider range of objects with the goal of identifying trends in the thermal structures and molecular abundances and how they correlate with empirical metrics. This will undoubtably verify hypothesized physical and chemical mechanisms operating in brown dwarf atmospheres, and likely identify unknown ones as well.

\section{Acknowledgements}	
The authors would like to thank Adam Burgasser, Brendan Bowler, Kelle Cruz, Mike Cushing, Michael Liu, and Emily Rice for useful discussions on benchmark systems, data treatment, and various data-model comparison approaches.  The authors thank Richard Freedman and Roxana Lupu for providing gas opacities and Caroline Morley for radiative transfer code comparisons and helpful discussions.  We thank Jacob Lustig-Yeager and Kyle Luther for re-writing portions of the code in python and C for significant speed improvements and also Dan Foreman-Mackey for making EMCEE available to the community.  Finally, we thank the anonymous referee and statistics consultant for useful and insightful comments. JT acknowledges financial support from the Carnegie Origins Postdoctoral Fellowship Program.  BB acknowledges financial support from the European Commission in the form of a Marie Curie International Outgoing Fellowship (PIOF-GA-2013-629435). JF acknowledges funding support from NSF award AST-1312545. MM acknowledges support from the NASA Astrophysics Theory and Planetary Atmospheres programs

\section{Appendix}\label{sec:appendix}
%Add a paragraph introduction--get Johanna to do it
Comparing the carbon-to-oxygen ratio in stellar-brown dwarf companion systems provides an additional benchmark dimension. Both brown dwarfs and the stars that they orbit are presumed to form out of the same molecular cloud, and thus would be expected to have the same elemental abundances. Metallicity and age are usually the benchmark dimensions.  Additional dimensions provide more constraints on the system properties.  Determining stellar abundances is no easy task as different groups using different techniques with different data on the same objects often times report significantly different results (Hinkle et al. 2014).  In this section we provide our own analysis to determine the stellar carbon-to-oxygen ratios in Gl570A and HD3651A.

\subsection{Observations and Stellar Parameter Analysis}
\subsubsection{Gl570A}
The observations of the K4 dwarf Gl570A were conducted on 13 July 2014 (UT) with the Magellan Inamori Kyocera Echelle (MIKE) spectrograph (Bernstein et al. 2003) on the 6.5m Landon Clay (Magellan II) Telescope at Las Campanas Observatory. Three frames of 100s each were taken of the target with the 0.5''x5'' slit and 1x1 binning. On 12 July 2014 (UT), three frames of 500s each with the 0.35''x5'' slit and 1x1 binning were taken of Vesta, as a solar standard. On both nights calibrations (biases, quartz and milky flats, and ThAr lamp spectra) were taken at the beginning of the night. MIKE is a double echelle spectrograph, meaning a dichroic splits the light into blue (3350-5000 {\AA} ) and red (5000-9400 {\AA}) arms. The data were reduced, extracted, combined, and wavelength calibrated with the Carnegie Python Distribution (CarPy) MIKE pipeline, written by D. Kelson (see also Kelson 2003). The resulting S/N in the Gl570A spectrum was $\sim$160, and $\sim$200 at 6300 {\AA} . Continuum normalization, order stitching, and Doppler-shift correction were performed with standard packages in IRAF\footnote{IRAF is distributed by the National Optical Astronomy  Observatory, which is operated by the Association of Universities   for Research in Astronomy, Inc., under cooperative agreement with the National Science Foundation.}.

Though the stellar parameters of Gl570A have been measured previously (e.g., Feltzing \& Gustafsson 1998; Thor{\'e}n \& Feltzing 2000; Ghezzi et al. 2010; Lee et al. 2011), for consistency we re-derived the T$_{\rm{eff}}$, log $g$, microturbulence ($\xi$), and [Fe/H] values from the MIKE spectra, based on the methods from our previous work (e.g., Teske et al. 2014). Briefly, Gl570A's stellar parameters were derived from equivalent width (EW) measurements of Fe I and Fe II. We used the iron line list of Tsantaki et al. (2013), optimized for cool stars (T$_{\rm{eff}} <$ 5000 K) by matching spectroscopic and infrared flux method (IRFM) temperatures. We forced zero correlation between [Fe I/H] and lower excitation potential ($\chi$) to set T$_{\rm{eff}}$, zero correlation between [Fe I/H] and reduced equivalent width [log(EW/$\lambda$)] to set $\xi$, and zero difference (within two decimal places) between [Fe I/H] and [Fe II/H] to set log $g$. The abundances of Fe were determined using the local thermodynamic equilibrium (LTE) spectral analysis code MOOG (Sneden 1973), with model atmospheres interpolated from the Kurucz ATLAS9 NOVER grids\footnote{See http://kurucz.harvard.edu/grids.html.}. Equivalent widths were measured in IRAF with the `splot' task, and abundances were normalized to the solar values as measured in our Vesta spectrum on a line-by-line basis. The log$N$(Fe) values for the Sun were determined with our Vesta spectrum and a solar Kurucz model with $T_{\rm{eff}}$=5777 K, log $g$=4.44 dex, [Fe/H]=0.00 dex, and $\xi$=1.38 km s$^{-1}$. In Gl570A, 89 Fe I and 10 Fe II lines were measured; the line properties and EWs are provided in Table \ref{tab:lines}. %Conservative uncertainties in the stellar parameters were calculated as in Teske et al. (2014). 

\begin{table*}[h] \footnotesize
\caption{Lines Measured, Equivalent Widths, and Abundances}
\centering
\vspace{5pt}
\label{tab:lines}
\begin{tabular}{lccccccccc}
\hline
\hline
\colhead{Ion} & \colhead{$\lambda$} & \colhead{$\chi$} & \colhead{log $gf$} & \colhead{EW$_{\odot}$} & \colhead{log$N_{\odot}$} & \colhead{EW$_{\rm{GL570A}}$} &  \colhead{log$N_{\rm{GL570A}}$} & \colhead{EW$_{\rm{HD3651A}}$} &  \colhead{log$N_{\rm{HD3651A}}$}\\ 
 \colhead{ } & \colhead{({\AA})} & \colhead{(eV)} & \colhead{} & \colhead{(m{\AA})} & \colhead{ } & \colhead{(m{\AA})} & \colhead{} & \colhead{(m{\AA})} & \colhead{}  \\
\hline
C I  &         5052.17 &7.685 &-1.24   &  33.3& 8.38   & \nodata  & \nodata &  25.0 & 8.75 \\
C I  &          5380.34& 7.695&-1.57  &    20.9& 8.44  &  \nodata   & \nodata &  15.1 & 8.81 \\
$[$C I$]$ & 8727.13 & 1.26 & -8.165 & 5.3 & 8.43 & 6.8 & 8.568 & \nodata &\nodata  \\  
$[$O I$]$ & 6300.30 & 0.00 & -9.717 & 5.4 (5.0) &{8.67\footnote{Abundance derived through equivalent width analysis.} (8.62$^{a}$)} &{7.9} & {8.59$^{a}$} & 6.6 & 8.82$^{a}$\\
{ } & {} & {} & {} & { }&{8.68\footnote{Abundance derived through synthesis analysis.}(8.61$^{b}$)} &{} & {8.54$^{b}$} & {} & {8.82$^{b}$} \\
O I & 7771.94 & 9.15 & 0.369 &71.5 (69.7) &{8.86\footnote{\,LTE abundance.} (8.83$^{c}$)} &12.3& 8.85$^{c}$&38.4& 9.14$^{c}$ \\
{ } & 7774.17 & 9.15 & 0.220 &61.5 (63.2) &{8.86$^{c}$ (8.88$^{c}$)} & 11.2 & 8.93$^{c}$&38.3& 9.28$^{c}$\\
{ } & 7775.39 & 9.15 & 0.001 & 49.0 (44.8) &{8.86$^{c}$ (8.78$^{c}$)} & 7.30 & 8.87$^{c}$ &25.1& 9.15$^{c}$\\
\hline
\end{tabular}
\tablecomments{The number abundances (log$N$) for HD3651A listed in this table are calculated as an example with
Allende Prieto et al. (2004) stellar parameters. Any value in parentheses refers to a HIRES solar measurement; all solar-normalized HD3651A abundances in the text are relative to HIRES solar measurements. This table is available in its entirety in a machine-readable form in the online journal. A portion is shown here for guidance regarding its form and content.}
\end{table*}

Our final parameters and errors for Gl570A are listed below (Table \ref{tab:params}, along with those of several other studies for comparison. The errors are calculated as in our previous work (Teske et al. 2013ab, 2014, 2015) --  the change in $T_{\rm{eff}}$ ($\xi$) required to cause a correlation coefficient $r$ between [Fe I/H] and $\chi$ ([Fe I/H] and reduced
EW) significant at the 1$\sigma$ level was adopted as the uncertainty in these
parameters. The uncertainty in log $g$ was calculated differently,
through an iterative process described in detail in Baubar \&
King\,(2010). Uncertainties in [Fe I/H] and [Fe II/H] are calculated from the quadratic sum of the
individual uncertainties in these abundances due to the derived
uncertainties in $T_{\rm{eff}}$, log $g$, and $\xi$, as well as the uncertainty
in the mean ($\sigma_{\mu}$\footnote{$\sigma_{\mu}=\sigma/\sqrt{N-1}$,
where $\sigma$ is the standard deviation of the derived abundances and
$N$ is the number of lines used to derive the abundance.}) of each
abundance. The uncertainty due to the stellar parameters is measured from the
sensitivity of the abundance to each parameter for
changes of $\pm150$ K in $T_{\rm{eff}}$, $\pm0.25$ dex in log $g$, and
$\pm0.30$ km s$^{-1}$ in $\xi$.  The uncertainty due to each
parameter is then the product of this sensitivity and the
corresponding parameter uncertainty. For the abundances determined through
spectral synthesis (e.g., from [O I], see below), models with this range of stellar parameters were
compared to the data and the elemental abundance adjusted to determine the best fit.

While our log $g$ value is moderately lower than other some other studies, it agrees within errors. As described below, these stellar parameter errors are propogated through the other abundance measurements.

%\begin{center}
\begin{table}[h!] \footnotesize
\caption{Gl570A Stellar Parameters}
\centering
\vspace{5pt}
\label{tab:params}
\begin{tabular} {|l | c | c | c | c | c |}
\hline
Parameter &                   this work     & Feltzing &  Thor{\'e}n    &  Ghezzi et al. (10)  &  Lee et al. (11) \\
{ }              &                                     & \& Gustafsson (98) & \& Feltzing (00) & {} & {} \\
\hline
T$_{\rm{eff}}$ (K)        &  4686 $\pm$47     &        4585    & 4585&       4799 $\pm$72       &   4615  \\    
\hline
log $g$ (cgs)           &  4.37 $\pm$0.27    &       4.70    & 4.58&    4.60 $\pm$0.16          &   4.36  \\    
\hline
$\xi$  (km s$^{-1}$) & 1.03 $\pm$0.16     &        1.0     &  1.0         &      0.77$\pm$0.08       &   \nodata  \\    
\hline
[Fe/H] (dex)              & -0.05$\pm$0.17    &       0.04     & 0.04     &      0.03$\pm$0.03             &   -0.05 \\    
\hline
\end{tabular}
\end{table}
%\end{center}

\subsubsection{HD3651A}
The bright ($V$=5.88) K0 dwarf HD3651A hosts a 0.2 M$_J$ planet (Fischer et al. 2003), and has thus been the target of many spectroscopic observations and stellar parameter analyses ($\sim$15, according to SIMBAD). Given the many previous stellar parameter analyses based on high-resolution, high-S/N data, we do not derive yet another set of stellar parameters here. Instead we use an archive HIRES spectrum (J. Johnson 2015, private communication), with S/N$\sim$200 at the [O I] 6300 {\AA} line, along with several reported sets of stellar parameters (Table \ref{tab:params2}) to verify the previously-measured carbon and oxygen abundances. We also assess realistic uncertainties on these measurements, as no formal uncertainties were previously published; our analysis is reported in the next section. The solar standard in this case is an archive HIRES spectrum of reflected light from the asteroid Vesta (A. Howard 2014, private communication), taken in the same configuration as the HD3651A spectrum.

\begin{table}[h!] \footnotesize
\caption{HD3651A Previously Measured Stellar Parameters}
\centering
\vspace{5pt}
\label{tab:params2}
\begin{tabular} {|l | c | c | c | c | c |}
\hline
Parameter &      Allende Prieto   et al. (04)      &  Delgado Mena et al. (10) & Petigura \&   Marcy (11)& Ram{\'{\i}}rez et al. (13)\\
\hline
T$_{\rm{eff}}$ (K)        &  5117$\pm$94     &      5173         &       5221     & 5303$\pm$63  \\    
\hline
log $g$ (cgs)           &  4.58 $\pm$0.04   &     4.37          &    4.45         & 4.56$\pm$0.03\\    
\hline
$\xi$  (km s$^{-1}$) & 0.93                      &       0.74         &   \nodata   & 0.64$\pm$0.12\\    
\hline
[Fe/H] (dex)              & 0.13                       &        0.12         &  0.16      & 0.18$\pm$0.07\\    
\hline
\end{tabular}
\end{table}

\subsection{C/O Ratio Analysis}

\subsubsection{Carbon Measurements}
\subsubsection{Gl570A}
The determination of [C/H] in Gl570A required a different technique than our previous work, due to the low temperature of the star. The carbon abundances for Gl570A derived from EW measurements of widely-used high-excitation ($\chi \ge$ 7.67 eV) C I lines (5380, 7711, 7113{\AA} ) and a line-by-line comparison to solar resulted in [C/H]$_{\rm{avg}}$=0.99, an unrealistically high value. These C I lines suffer NLTE effects, such that an LTE analysis overestimates the abundances, and corrections based on NLTE atomic models are predicted to increase in magnitude (larger negative corrections) with higher T$_{\rm{eff}}$ and lower log $g$ (e.g., Asplund 2005; Takeda\& Honda 2005; Fabbian et al. 2006). However, similar to the O I triplet lines at 7771/7774/7775{\AA}  (see discussion below), the NLTE corrections for cool stars are predicted to be minimal, less than the solar-type star corrections of $\sim$-0.05. As discussed in Teske et al. (2013a) for the cool star ($\sim$5350 K) 55 Cnc, applying the predicted NLTE corrections to the O I triplet abundances actually \textit{increases} the [O/H] values, rather than decreases, because the solar corrections exceed the cooler-star corrections and thus their differences are increased. We find the same problem with the C I lines in Gl570A.

Thus, the [C/H] for Gl570A is instead derived from the low-excitation ($\chi$=1.26 eV) forbidden [C I] line at 8727.13{\AA} (Figure \ref{fig:lines}). This line is weak, and possibly blended with an Fe I line at 8727.10{\AA}  (Lambert \& Swing 1967), but has a well-determined transition probability and is not susceptible to depatures from LTE (Gustafsson et al. 1999; Asplund et al. 2005). Using a spectral synthesis analysis with a line list between 8724-8730{\AA} , gathered from the Vienna Atomic Line Database (VALD; Kupka et al. 1999), we derive A(C)$_{\rm{GL570A}}$=8.57. This also agrees with our derivation using the IRAF-measured EW (5.30 m{\AA} ) and the `abfind' driver in MOOG with the Gl570A model derived from the same MIKE spectra. Via the same procedure, we measure A(C)$_{\rm{solar}}$=8.43 from the 8727 {\AA}  [C I] line. The resulting [C/H] is listed in Table \ref{tab:abuns1}. Note that in Table \ref{tab:abuns1}, we include the \textit{formal} errors on each abundance measurement ([Ni/H], [C/H], [O/H]), which is the quadratic sum of the three individual parameter uncertainties ($T_{\rm{eff}}$, log $g$, $\xi$) and $\sigma_{\mu}$ as described above for the case of iron.

\subsubsection{HD3651A}
The [C I] line used to measure the carbon abundance of Gl570A was not available in our HD3651A HIRES spectrum, so we instead rely on the lowest excitation ($\chi$=7.685 eV) available C I lines at 5052.2 {\AA} and 5380.3 {\AA}. These lines, as mentioned above, are predicted to have negligible NLTE corrections at the low temperature of HD3651A ($\sim$5100-5200 K), on par with those predicted for the Sun ($\le$0.05 dex; Takeda \& Honda 2005). Thus, any deviations from LTE should cancel with the calculation of the solar-normalized [C/H] value derived from these lines. In this case, the C I lines resulted in reasonable [C/H] values (see Tables \ref{tab:lines} and \ref{tab:abuns2}). Measurements of other available C I lines used in previous work (6588, 7111, 7113{\AA}, e.g. Teske et al. 2014) result in $\sim$0.1-0.25 dex larger [C/H] values -- as expected since these higher excitation lines are more susceptible to NLTE effects -- and thus are excluded for the final [C/H] for HD3651A. We use our equivalent width measurements (Table \ref{tab:lines}) with three different stellar models (Allende Prieto et al.'s, Delgado Mena et al.'s, and Petigura \& Marcy's, see Table \ref{tab:params2}) to derive carbon abundances, and use the resulting spread in [C/H] as a measure of its uncertainty. Our measured [C/H] values (0.21-0.37) agree with those of Allende Prieto et al. (2004) and Delgado Mena et al. (2010), as expected (Table \ref{tab:abuns2}). Note that in Table \ref{tab:abuns2}, we include only the \textit{standard deviation} ($\sigma$) errors, since the stellar parameter errors are not derived in this work, and are calculated differently for each source; the error calculation method used for Gl570A does not apply. 

\subsubsection{Oxygen Measurements}
As discussed in Teske et al. (2013) and (2014), and by many others in the past (e.g., Nissen \& Edvardsson 1992; King \& Boesgaard 1995; Asplund et al. 2004; Schuler et al. 2006a,b; Caffau et al. 2008), oxygen abundances are notoriously difficult to derive, particularly in stars that are cooler/hotter and/or more/less metal-rich than the Sun. The oxygen abundance indicators we explored here include the forbidden [O I] line at 6300{\AA}, which is well-described by LTE but blended with a Ni I line, and the O I triplet at 7771-7775{\AA} , made up of three unblended and usually prominent lines amenable to direct EW measurement. 

\begin{figure}[ht!]
\includegraphics[width=1.0\textwidth]{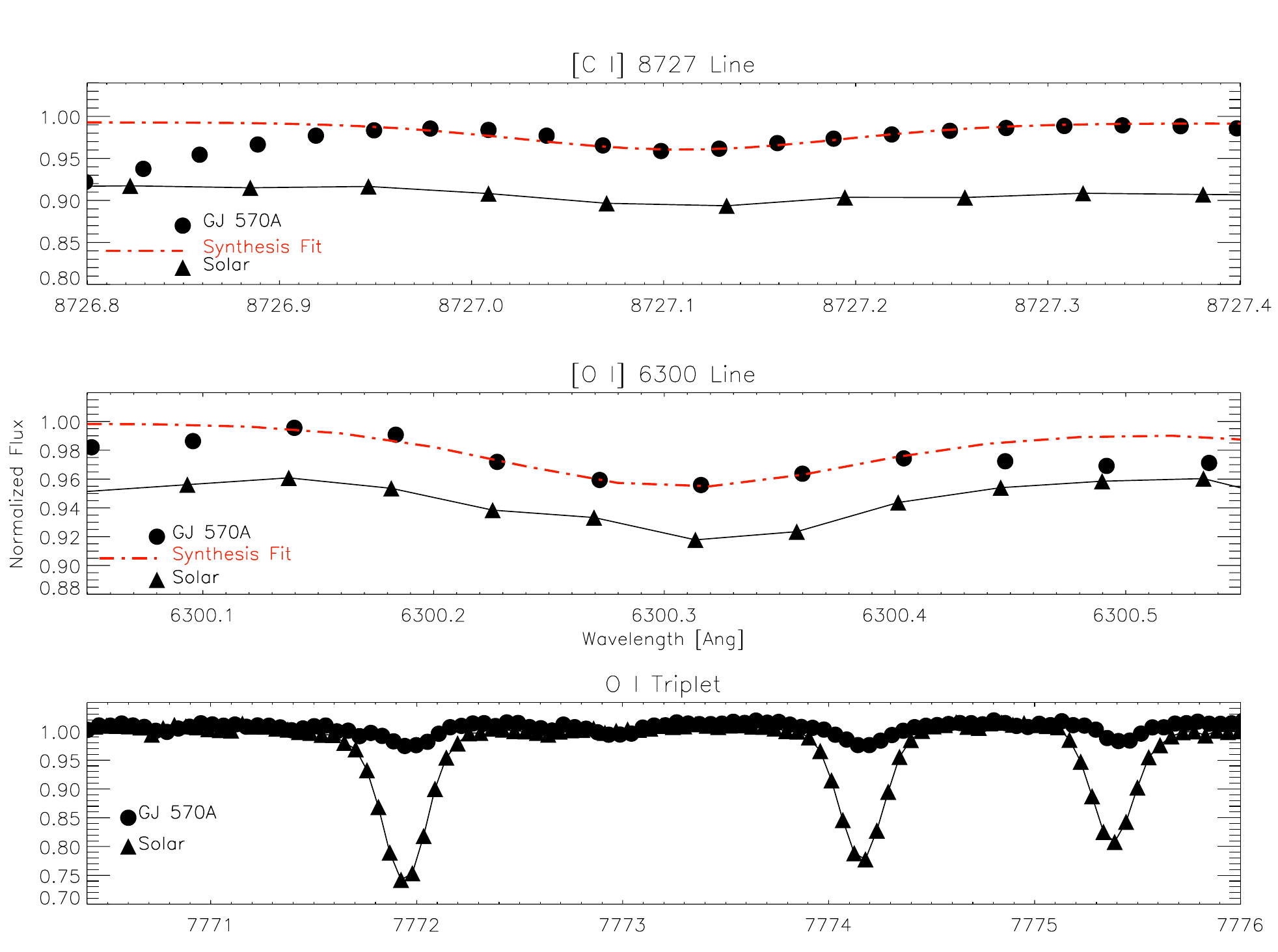}
 \caption{Plotted are the the lines measured in this work to determine the C/O ratio of Gl570A. Dots represent the spectrum of Gl570A, while triangles represent the solar standard spectrum. In the top two plots, synthesis fits to the lines are shown in red dot-dashed lines. The result abundances are given in Table \ref{tab:abuns1}.}
\label{fig:lines}
\end{figure}

We derived the [O I] 6300.304{\AA}  line using two methods. First, we measured the EW of the feature directly from the spectrum as input for the `blends' driver in MOOG, accounting for $^{60}$Ni I $+$ $^{58}$Ni I feature at 6300.335{\AA}  with log$gf$($^{60}$Ni)=-2.695 and log$gf$($^{58}$Ni)=-2.275 as derived by Bensby et al. (2004). This resulted in [O/H]$_{\rm{6300,blends}}=$ -0.08 for Gl570A and 0.19-0.23 for HD3651A, depending on the set of stellar parameters. Second, we performed a spectral synthesis analysis on the line, using as a ``known'' our measured [Ni/H] abundance based on a line-by-line analysis with the Sun (as with Fe; see Tables \ref{tab:abuns1} and \ref{tab:abuns2}). The synthesized spectra (with the stellar parameters derived or noted above) were convolved with a Gaussian profile, based on near-by unblended lines, to represent the instrument PSF, stellar macroturbulence, and rotational broadening; we also fixed the nickel abundance to our measured value. Unlike in the case of 55 Cnc (Teske et al. 2013a), the oxygen line strength did not change drastically ($\lesssim$0.02 dex) by changing the Ni abundance within our derived [Ni/H]. The remaining free parameters were continuum normalization, wavelength shift, and oxygen abundance. The best fit to the synthesized spectra for the [O I] was determined by minimizing the deviations between the observed and synthetic spectra (see Figure \ref{fig:lines}). This resulted in [O/H]$_{\rm{6300,synth}}=$ -0.13 for Gl570A and 0.17-0.27 for HD3651A, depending on the set of stellar parameters. For our final [O/H]$_{6300}$ we took the mean of these measurements, -0.11$\pm$0.19 dex for Gl570A and 0.18-0.25 dex for HD3651A.

The O I triplet suffers NLTE effects due to the dilution of the each line's source function with respect to the Planck function (e.g., Kiselman 1993; Gratton et al. 1999; Kiselman 2001), so abundances derived assuming LTE are overestimated. As with C I, the predicted NLTE corrections increase with decreasing gas pressures and/or increasing temperatures, but decrease for cool stars (e.g., Takeda 2003; Ram{\'{\i}}rez et al. 2007, Fabbian et al. 2009) like both Gl570A and HD3651A. We measured the EWs of the three O I triplet lines and derived an A(O) for each line in Gl570A, HD3651A, and their respective solar standards (see Table \ref{tab:lines}); the average [O/H]$_{\rm{LTE}}$ of the line-by-line differences with the Sun is 0.05$\pm$13 for Gl570A (line 3 of Table \ref{tab:abuns1}) and 0.17-0.38 for HD3651 (lines 3, 6, and 9 of Table \ref{tab:abuns2}). We then applied NLTE corrections from three different sources  -- Takeda (2003), Ram{\'{\i}}rez et al. (2007), and Fabbian et al. (2009) -- to Gl570A, HD3651, and the respective solar standard measurements, and recalculated the line-by-line abundance differences (see discussion in Teske et al. 2013a for details of each of these correction schemes). Corrections in absolute abundance (not relative to solar) for the Sun range from 0.13-0.21, for Gl570A range from 0.04-0.07, and for HD3651 range from 0.52-0.85. The resulting [O/H]$_{\rm{NLTE}}$, averaged over all three lines and all three sources of NLTE corrections, is 0.14 dex for Gl570A (line 4 of Table \ref{tab:abuns1}) and 0.25-0.46 dex, depending on the stellar parameters. In the case of Gl570A, based on previous NLTE abundance uncertainty calculations, we assume the same uncertainty as the LTE abundance; for HD3651, again the $\sigma$ (standard deviations) across the three lines and three sources are listed in Table \ref{tab:abuns2}. 

\begin{table}[h!] \footnotesize
\caption{Gl570A Abundances \& Indicators with Formal Errors}
\centering
\label{tab:abuns1}
\vspace{5pt}
\begin{tabular} {|l | c | c | c | c | c |}
\hline
Source          &[Ni/H] (dex) & [C/H] (dex) & [O/H] (dex) & [O/H] indicator & C/O      \\
\hline
this work     &  0.01 $\pm$0.05            & 0.14$\pm$0.11  &   -0.11$\pm$0.19  &[O I] 6300 {\AA}        & 0.97$\pm$0.22 \\
 {    }              & 0.01 $\pm$0.05            &0.14$\pm$0.11  & 0.05$\pm$0.13          &  O I triplet 7775 {\AA}   LTE    &  0.68$\pm$0.17\\
 {    }             & 0.01 $\pm$0.05            & 0.14$\pm$0.11 &   0.14$\pm$0.13          & O I triplet 7775 {\AA}    NLTE    & 0.55$\pm$0.17 \\
\hline
Feltzing \& Gustafsson (98) &0.16          & 0.18 $\pm$0.18            & \nodata       &[O I] 6300 {\AA}          & \nodata \\
\hline
Petigura \& Marcy  (11)       &  0.15 $\pm$0.06         & 0.18            & \nodata       &[O I] 6300 {\AA}          & \nodata \\
\hline
\end{tabular}
\end{table}

\begin{table}[h!] \footnotesize
\caption{HD3651A Abundances \& Indicators with $\sigma$ Errors}
\centering
\label{tab:abuns2}
\vspace{5pt}
\begin{tabular} {|l | c | c | c | c | c |}
\hline
Source                         &[Ni/H] (dex) & [C/H] (dex) & [O/H] (dex) & [O/H] indicator & C/O      \\
\hline
this work,                         &  0.19$\pm$0.04            & 0.37$\pm$0.002  &   0.21$\pm$0.01       &[O I] 6300 {\AA}        & 0.79 \\
  AP04 Params                      & 0.19$\pm$0.04                 &0.37$\pm$0.002     & 0.36$\pm$0.05         &  O I triplet 7775 {\AA}   LTE    &  0.56\\
 {    }                              & 0.19$\pm$0.04                 & 0.37$\pm$0.002     &   0.44$\pm$0.05         & O I triplet 7775 {\AA}    NLTE    & 0.47 \\
\hline
this work,                         &  0.22 $\pm$0.05            & 0.25$\pm$0.004  &   0.18$\pm$0.02        &[O I] 6300 {\AA}        & 0.65 \\
DM10 Params                  & 0.22 $\pm$0.05            &0.25$\pm$0.004     & 0.24$\pm$0.05          &  O I triplet 7775 {\AA}   LTE    &  0.56 \\
 {    }                              & 0.22 $\pm$0.05            & 0.25$\pm$0.004      &   0.31$\pm$0.05        & O I triplet 7775 {\AA}    NLTE    &    0.48    \\
\hline
this work,                          &  0.19 $\pm$0.04            & 0.21$\pm$0.001  &   0.25$\pm$0.03         &[O I] 6300 {\AA}        & 0.50 \\
PM11 Params                      & 0.19 $\pm$0.04            &0.21$\pm$0.001    & 0.16$\pm$0.05          &  O I triplet 7775 {\AA}   LTE    &  0.62\\
 {    }                               & 0.19 $\pm$0.04           & 0.21$\pm$0.001      & 0.23$\pm$0.04        & O I triplet 7775 {\AA}    NLTE    &        0.52  \\
\hline
\hline
\hline
Allende Prieto et al. (04) & 0.27                             & 0.26                 &     0.23                           &[O I] 6300 {\AA}          & 0.59  \\
\hline
Delgado Mena et al. (10) & 0.15                              & 0.25                     & 0.06                        &[O I] 6300 {\AA}          & 0.85 \\
\hline
Petigura \& Marcy  (11)       &  0.24                         & \nodata               &0.07$\pm$0.08                         &[O I] 6300 {\AA}          & \nodata \\
\hline
Ram{\'{\i}}rez et al. (13)      &  \nodata                        & \nodata         &  0.05$\pm$ 0.04   &O I triplet 7771-5 {\AA}  LTE        & \nodata \\
{}                                        &  \nodata                        & \nodata         &  0.12$\pm$0.04   &O I triplet 7771-5 {\AA}   NLTE       & \nodata \\
\hline
\end{tabular}
\end{table}

\subsection{What are the C/O Ratios of Gl570A and HD3651A?}

We calculate the C/O ratio\footnote{The C/O ratio -- the ratio of the number of carbon atoms
   to oxygen atoms -- is calculated in stellar abundance analysis as C/O=
  N$_{\rm{C}}$/N$_{\rm{O}}$=10$^{\rm{logN(C)}}/10^{\rm{logN(O)}}$ where
 log(N$_{\rm{X}}$)=log$_{10}$(N$_{\rm{X}}$/N$_{\rm{H}}$)+12.} of Gl570A and HD3651A with the Asplund et al. (2009) solar A(C)=8.43 and A(O)=8.69 values and our [C/H] and [O/H] values as follows:

\begin{center}

C/O $=$ 10$^{8.43+[C/H]}/10^{8.69+[O/H]}$

\end{center} 

\noindent with the error on C/O represented by the errors of [C/H] and [O/H] added in quadrature. From our different [O/H] indicators, the C/O ratio for Gl570A ranges from 0.55$\pm$0.17 to 0.97 $\pm$0.22, and for HD3651A ranges from 0.47-0.79, depending on the set of stellar parameters and oxygen abundance indicators. 

The [O I] line has been designated as a consistently reliable oxygen abundance indicator (e.g., Lambert 1978; Allende Prieto et al. 2001; Asplund et al. 2004; Schuler et al. 2006a) due to its formation in the ground state making an LTE approximation exceedingly good (Caffau et al. 2008). As noted above, the O I triplet suffers significant NLTE effects, which are predicted to decrease with temperature. However, studies by Schuler et al.\,(2004,2006b) and King \& Schuler (2005) of dwarf stars in the Pleiades, M34, and Hyades open clusters, and the Ursa Major moving group found that [O/H]$_{\rm{triplet,LTE}}$ values derived from the O I triplet significantly \textit{increased} with decreasing $T_{\rm{eff}}$ ($\lesssim$5400). If the assumption holds that stars within a single cluster or moving group should be chemically homogenous, the increasing [O/H]$_{\rm{triplet,LTE}}$ with $T_{\rm{eff}}$ is in direct contrast with all the available NLTE calculations. These studies do not point to a definitive cause of the NLTE correction discrepancy in cool dwarfs, though they suggest age likely plays a role. However, the studies do indicate for cool stars like Gl570A and HD3651A, the O I triplet-tempearture trend appears to contradict the predicted NLTE oxygen abundance corrections. Due to larger corrections for the Sun versus Gl570A and HD3651A, applying the NLTE-corrections of  Takeda (2003), Ram{\'{\i}}rez et al. (2007), and Fabbian et al. (2009) all result in [O/H] values for Gl570A and HD3651 that are in general \textit{larger} than in the LTE case, and C/O values that are in general \textit{smaller}. 

Given the caveats of the O I triplet NLTE values, we chose here to combine the [O/H]$_{\rm{triplet,LTE}}$ values from the three O I lines, and [O/H]$_{6300}$ for our final best estimate for the oxygen abundances of Gl570A and HD3651A, resulting in [O/H]$_{\rm{avg}}=$ -0.03$\pm$0.12 for Gl570A (where the uncertainty here is the errors of the O I and [O I] abundances added in quadrature) and [O/H]$_{\rm{avg}}=$0.21-0.29 for HD3651A, depending on the set of stellar parameters. With [C/H]$_{\rm{GL570A}}=$0.14$\pm$0.11, our final C/O for Gl570A $=$0.81$\pm$0.16. For HD3651, taking [O/H]$_{\rm{avg}}=$0.23, $\sigma=$0.07 and [C/H]$_{\rm{avg}}=$0.28, $\sigma=$0.08 from the three different stellar parameter analyses, the final C/O$_{\rm{avg}}$ for HD3651A=0.62, $\sigma=$0.11. 

\subsection{Comparison with the Literature}
\subsubsection{Gl570A}
Our [O/H]$_{\rm{avg}}$ differs from the [O/H] values for Gl570A reported by Feltzing \& Gustafsson (1998) and Petigura \& Marcy (2011), who both find [O/H]$\sim$0.15. This higher oxygen abundance, combined with our measured [C/H] (none of the other studies measured carbon in Gl570A), would lower the C/O ratio to 0.54, which matches the solar value (C/O$_{\odot}$=0.55$\pm$0.10; Asplund et al.\,2009; Caffau et al.\,2011). Considering our errors, and those reported by Petigura \& Marcy (2011) for [O/H], the high and low C/O values would just barely overlap within errors. 

Both Feltzing \& Gustafsson (1998) and Petigura \& Marcy (2011) also measure higher [Ni/H] values (0.18 dex), although the Feltzing \& Gustafsson abundance overlaps with ours within errors. Using a stellar model with Petigura \& Marcy (2011)'s derived parameters for Gl570A (4744 K $T_{\rm{eff}}$, 4.76 dex log $g$, 0.10 dex [Fe/H], and assuming $\xi=$1.00 km s$^{-1}$) and our measured EWs for Ni I, C I, [O I], and the O I triplet in LTE results in [Ni/H]$=$0.14 and C/O$=$10$^{8.43+.37}/10^{8.69+(0.12+0.20)/2}=$0.89. Performing the same exercise with Feltzing \& Gustafsson (1998)'s derived stellar parameters (4585 K T$_{\rm{eff}}$, 4.70 dex log $g$, 0.04 dex [Fe/H], 1.00 km s$^{-1} \xi$) results in [Ni/H]$=$0.14 and C/O$=$10$^{8.43+.33}/10^{8.69+(0.40+0.14)/2}=$0.63, where in both formulae the average [O/H]$=$([O/H]$_{\rm{triplet,LTE}} +$[O/H]$_{6300}$)/2. 
%PM C 0.368 O triplet 0.12 O forb 0.196
%FG C   .33    O 0.40      O 0.136
Given our uncertainties and the typical uncertainties in [C/H] and [O/H], particularly in cool stars, these alternative C/O values and our reported C/O agree within errors. This exercise also demonstrates that the different [O/H]$_{6300}$ and [Ni/H] abundances derived in this work versus in Feltzing \& Gustafsson (1998) and Petigura \& Marcy (2011) are likely due to differences in stellar parameters, and not the quality of the spectra or the empirical measurements -- using our measurements and their models results in [O/H]$_{6300}$ and [Ni/H] values very similar to what they report. 

%\textit{Discuss agreement with Nissen et al. 2014 C/O relation.}

\subsubsection{HD3651A}
The [C/H] values we derive for HD3651A based on our measurements of a high-S/N archive HIRES spectrum are in decent agreement with those reported by Allende Prieto et al. (2004) and Delgado Mena et al. (2010). Since our [C/H] is derived from stronger C I lines, versus Allende Prieto's [C I] 8727 {\AA} measured [C/H], we might expect our abundance to be higher, which it is. We have perfect agreement with Delgado Mena et al. (2010) when using their stellar parameters and our C I equivalent width measurements ([C/H]=0.25). 

The most challenging aspect of the C/O measurement in this (and many) stars is pinning down the [O/H]. In comparison to Allende Prieto et al. (2004)'s [O/H] derived from the [O I] 6300 {\AA} line, our [O I] 6300 {\AA} measurement combined with their stellar parameters produces almost an almost identical [O/H] (0.21 versus 0.23). This is not the case when comparing [O/H]$_{\rm{6300}}$ values of Delgado Mena et al. (2010) and Petigura \& Marcy (2011) to those measured here, where we find oxygen abundance higher by 0.12 and 0.18, respectively. Other authors (Fortney 2012; Nissen 2013; Teske et al. 2013a; Nissen et al. 2014) have called into question the high C/O ratios (often caused by low oxygen abundances) reported in the previous papers, and in some cases have reported different [O/H] results for the same stars. Differences in [O/H] measured from the same line in the same star could arise due to several challenging aspects of the 6300 {\AA} line, such as continuum placement, telluric contamination, weakness of the line at low metallicity, and blending with an Ni I line that can make up to 30\% of the line strength in the Sun (Caffau et al. 2008). Delgado Mena et al. (2010) specifically removed spectra with obvious telluric contamination, and estimated the EW of the Ni line blended with [O I] to form the 6300 {\AA} line using the ``ewfind'' driver of MOOG and the Ni abundances measured from $\sim$50 Ni I lines. The [Ni/H] value we derived (0.22) from 27 Ni I lines and the same stellar parameters at Delgado Mena et al. is slightly lower than their value (0.15). Presumably a smaller Ni abundance comes from a smaller Ni EW, meaning a smaller contribution to the 6300 {\AA} line and thus a larger contribution from O, and yet Delgado Mena et al. also find a smaller [O/H]. The Ni I and O I line parameters that we use (Bensby et al. 2004 for Ni I; Story \& Zeippen 2000 for [O I]) differ from those used by Delgado Mena et al. (Allende Prieto et al. 2001, for Ni I; Lambert et al. 1978 for [O I]), but this is not enough to account for the 0.12 dex difference. Without a published EW measurements or synthesis fits, it is unclear why our [O/H]$_{\rm{6300}}$ for HD3651A differs from Delgado Mena et al.'s. 

In their [O I] 6300 {\AA} measurements, Petigura \& Marcy (2011) also discard any stars with telluric line contamination, but many of their spectra are contaminated by iodine lines, which are $\sim$5\% deep in the region of the line. To account for this, they shift the stellar spectrum (with iodine) to match in wavelength the most recent iodine reference spectrum, and then divide the stellar spectrum by the iodine reference spectrum. The resulting spectrum can contain artifacts at the $\sim$1\% level, likely within any EW measurement. Petigura \& Marcy treat the Ni blend differently: Once they derive an [O/H] from the 6300 {\AA} line by comparing synthetic spectra to their observed spectra, they refit the oxygen line to spectra with $\pm$0.03 dex Ni, and add the resulting errors in [O/H] in quadrature to their statistical errors. Petigura \& Marcy authors also use different line parameters for the Ni I line blended with oxygen at 6300 {\AA}), determined by fitting their solar spectrum to match their adopted solar abundance distribution. The [Ni/H] abundance derived by Petigura \& Marcy for HD3651A (0.24) agrees with what we derive using our EWs measured from 27 Ni I lines and their stellar parameters, but they do not explicitly use the measured nickel abundance for each star in their measurement of [O/H]$_{\rm{6300}}$. This may be the reason behind our differing  [O/H]$_{\rm{6300}}$ values. 

Ram{\'{\i}}rez et al. (2013) do not measure [O/H] from the 6300 {\AA} forbidden line, but instead from the O I 7771-5 {\AA} triplet. As noted above, these lines suffer NLTE effects that are not well understood or calibrated for cool stars. Combining our triplet line EWs with the stellar parameters of HD3651A from Ram{\'{\i}}rez et al. (who do not list their measured EWs) results in [O/H]$_{\rm{triplet,LTE}}$=0.11, $\sigma$=0.05, which is slightly higher than, but overlaps within errors of, Ram{\'{\i}}rez et al.'s [O/H]$_{\rm{triplet,LTE}}$=0.05$\pm$0.04 (where here we have added Ram{\'{\i}}rez et al.'s  line-to-line scatter, 0.01, and their uncertainty in the stellar parameters, 0.04, in quadrature for a total error). Similarly, we find [O/H]$_{\rm{triplet,NLTE}}$=0.19, $\sigma$=0.04, using our triplet EWs, Ram{\'{\i}}rez et al.'s stellar parameters and Ram{\'{\i}}rez et al. (2007)'s NLTE correction scheme, whereas Ram{\'{\i}}rez et al. (2013) reports [O/H]$_{\rm{triplet,LTE}}$=0.12$\pm$0.04 for HD3651A. Thus, while slightly higher, our [O/H] values still consistent with Ram{\'{\i}}rez et al.'s within errors; if we use [O/H]$_{\rm{triplet,LTE}}$ based on Ram{\'{\i}}rez et al.'s stellar parameters, and recalculate [C/H] with the same parameters, the resulting HD3651 C/O=0.66, in good agreement with our final value above (0.62$\pm$0.11).

{\it Facilities:} \facility{Magellan:Clay}

\end{document}